\documentclass[twocolumn]{aastex631}

\usepackage{chemformula}
\let\ce\ch
\usepackage{graphicx}
\usepackage{wasysym}
\usepackage{rotating}
\usepackage{textcomp}
\usepackage{array}

\begin{document}

\title{Constraints on the Size and Composition of the Ancient Martian Atmosphere from Coupled \ce{CO2}-\ce{N2}-\ce{Ar} Isotopic Evolution Models}

\author[0000-0003-2457-2890]{Trent B. Thomas}
\affiliation{Jet Propulsion Laboratory, California Institute of Technology, Pasadena, CA, USA}
\affiliation{Department of Earth and Space Sciences/Astrobiology Program, University of Washington, Seattle, WA, USA}

\author[0000-0003-2215-8485]{Renyu Hu}
\affiliation{Jet Propulsion Laboratory, California Institute of Technology, Pasadena, CA, USA}
\affiliation{Division of Geological and Planetary Sciences, California Institute of Technology, Pasadena, CA, USA}

\author[0000-0002-9603-500X]{Daniel Y. Lo}
\affiliation{Department of Climate and Space Sciences and Engineering, University of Michigan, Ann Arbor, MI, USA}

\begin{abstract}

Present-day Mars is cold and dry, but mineralogical and morphological evidence shows that liquid-water existed on the surface of ancient Mars. In order to explain this evidence and assess ancient Mars's habitability, one must understand the size and composition of the ancient atmosphere. Here we place constraints on the ancient Martian atmosphere by modeling the coupled, self-consistent evolution of atmospheric \ce{CO2}, \ce{N2}, and \ce{Ar} on Mars from 3.8 billion years ago (Ga) to the present. Our model traces the evolution of these species' abundances and isotopic composition caused by atmospheric escape, volcanic outgassing, and crustal interaction. Using a Markov-Chain Monte Carlo method to explore a plausible range of parameters, we find hundreds of thousands of model solutions that recreate the modern Martian atmosphere. These solutions indicate that Mars's atmosphere contained 0.3-1.5 bar \ce{CO2} and 0.1-0.5 bar \ce{N2} at 3.8 Ga. The global volume of deposited carbonates critically determines the ancient atmospheric composition. For example, a $\sim$1 bar \ce{CO2} ancient atmosphere with 0.2-0.4 bar \ce{N2} requires $\sim$0.9 bar \ce{CO2} deposited in carbonates primarily in open-water systems. With the joint analysis of \ce{C}, \ce{N}, and \ce{Ar} isotopes, we refine the constraints on the relative strengths of outgassing and sputtering, leading to an indication of a reduced early mantle from which the outgassing is sourced. Our results indicate that a \ce{CO2}-\ce{N2} atmosphere with a potential \ce{H2} component on ancient Mars is consistent with Mars's geochemical evolution and may explain the evidence for its past warm and wet climate.

\end{abstract}

\section{Introduction} \label{sec:intro}

Evidence for liquid-water on ancient Mars's surface has yet to be reconciled with our knowledge of ancient Mars's atmosphere and climate. Modern Mars is cold and has a thin atmosphere, leaving liquid-water unstable to evaporation or freezing on most of its surface \citep{haberle_possibility_2001}. However, geomorphological and mineralogical evidence convincingly shows abundant liquid-water existed at least transiently on Mars's surface 3 billion years ago (Ga) and earlier \citep[e.g.,][]{fassett_sequence_2011}. Thus, the climate on ancient Mars must have been significantly different from today. A thicker ancient atmosphere with various additional components has been proposed to explain this evidence, but the size, composition, and warming mechanism remain unknown empirically \citep[e.g.,][]{wordsworth_climate_2016}. The putative ancient atmosphere must be consistent not only with the evidence for liquid-water, but also with Mars's geochemical evolution. In other words, if there was a thick, multi-component atmosphere on ancient Mars, then it must have evolved to have the modern size and composition. Reconciling these aspects of ancient Mars will allow more precise evaluation of Mars's biological potential and a better understanding of how Mars's surface environment has changed over time.

\ce{CO2} alone cannot provide the greenhouse warming necessary to explain the evidence for liquid-water, but the addition of \ce{N2} and \ce{H2} may help. 1-dimensional and 3-dimensional climate models show that an atmosphere containing only \ce{CO2} and \ce{H2O} cannot warm ancient Mars enough to explain the geologic evidence, regardless of the atmosphere's size \citep{kasting_co2_1991,forget_3d_2013,wordsworth_global_2013}. One or more secondary atmospheric greenhouse gases are likely required. To this end, \ce{H2} is a potential component of the ancient atmosphere that can cause at least episodic warming of the surface via \ce{CO2}-\ce{H2} collision induced absorption (CIA) \citep{ramirez_warming_2014,wordsworth_transient_2017,ramirez_climate_2020,wordsworth_coupled_2021}. Additionally, isotopic evolution models indicate that the ancient atmosphere may have contained substantial \ce{N2} \citep{hu_nitrogen-rich_2022}, which can also contribute to surface warming. \citet{von_paris_n2-associated_2013} show that up to 13K warming on ancient Mars could come from \ce{N2}-\ce{N2} CIA and pressure broadening of \ce{CO2} absorption lines, but potential warming from \ce{H2} was not included. When combined with \ce{H2}, \ce{N2}-\ce{H2} CIA is a powerful greenhouse mechanism even with small amounts of \ce{H2} \citep{wordsworth_hydrogen-nitrogen_2013}. Warming from these mechanisms may also be enhanced by high-altitude clouds \citep{urata_simulations_2013, ramirez_could_2017, kite_warm_2021}. From a climate perspective alone, various combinations of these mechanisms may provide enough warming on ancient Mars to be consistent with the geologic evidence for water. 

As an alternative to reverse engineering the ancient atmosphere from the evidence for liquid-water, knowledge of Mars's geochemical evolution can be used to constrain the ancient atmosphere. Specifically, the isotopic composition of present-day Mars offers a window into how the atmosphere has evolved. Planetary processes occurring throughout Mars's history leave distinct fingerprints on the present-day atmosphere's isotopic composition \citep{jakosky_mars_1991}. For example, thermal and non-thermal atmospheric escape processes on Mars preferentially eject the lighter isotope of a given atmospheric species, causing isotopic fractionation. The different modes of atmospheric escape (e.g., sputtering, photochemical reactions) fractionate Mars's atmosphere at different rates. Deposition and sequestration of volatiles as minerals (e.g., carbonates and nitrates) and organics also fractionates the atmosphere, with preference to either the light or heavy isotope \citep{faure_principles_1991, lammer_loss_2020, house_depleted_2022}. Volcanically outgassed species introduced to the atmosphere will have an isotopic composition that reflects Mars's interior \citep[e.g.,][]{wright_chassigny_1992,mathew_early_2001}. Ultimately, the isotopic composition of present-day Mars's atmosphere is a product of these and other processes operating throughout Mars's history. By quantifying the rate at which they add or remove species from the atmosphere and their fractionation effect, it is possible to model the evolution of the mass and isotopic composition of Mars's atmosphere.

Isotopic evolution modeling studies place constraints on Mars's ancient atmospheric composition and its subsequent evolution \citep{jakosky_mars_1991,pepin_origin_1991,zahnle_xenological_1993,jakosky_mars_1994,pepin_evolution_1994,jakosky_history_1997,lammer_estimation_2003,kurokawa_evolution_2014,hu_tracing_2015,villanueva_strong_2015,kurokawa_interactive_2016,slipski_argon_2016,kurokawa_lower_2018, kurokawa_mars_2021,scheller_long-term_2021,hu_nitrogen-rich_2022}. An effective approach has been to find valid trajectories for Mars's atmospheric composition and evolution by comparing isotopic evolution models to present-day measurements of Mars's atmosphere. \citet{jakosky_mars_1994} and \citet{pepin_evolution_1994} laid the groundwork for modeling studies of this type, but more recent studies have been able to incorporate measurements from missions such as the Curiosity rover \citep{wong_isotopes_2013,atreya_primordial_2013, webster_isotope_2013} and MAVEN \citep{jakosky_mars_2015}. For example, \citet{hu_tracing_2015} model the evolution of \ce{CO2} and suggest the atmosphere was less than 1 bar at 3.8 Ga unless there was significant carbonate deposition in open-water systems. \citet{slipski_argon_2016} model the evolution of argon and place constraints on the rates of sputtering, volcanic outgassing, and crustal erosion. \citet{kurokawa_lower_2018} model the evolution of nitrogen and noble gases and find the atmosphere must be more than 0.5 bar at 4 Ga. \citet{kurokawa_mars_2021} model the evolution of neon and suggest there was recent active volcanism. \citet{scheller_long-term_2021} model the evolution of D/H in water and suggest that 30-99$\%$ of the water on ancient Mars was sequestered into minerals in the crust. \citet{hu_nitrogen-rich_2022} model the evolution of nitrogen and suggest that an atmosphere with hundreds of mbar \ce{N2} at 3.8 Ga is likely. These studies reveal the power of isotopic evolution modeling to constrain the ancient Martian atmosphere.

We present a comprehensive, coupled model for the isotopic evolution of \ce{CO2}, \ce{N2}, and \ce{Ar} in the Martian atmosphere, which are the three most abundant species today. This is the first study to self-consistently model the evolution of both the total abundances of \ce{CO2}, \ce{N2}, and \ce{Ar}, and their relative amounts of $\ce{^{12}C}$, $\ce{^{13}C}$, $\ce{^{14}N}$, $\ce{^{15}N}$, $\ce{^{36}Ar}$, $\ce{^{38}Ar}$, and $\ce{^{40}Ar}$. Few previous models include all three of \ce{CO2}, \ce{N2}, and \ce{Ar}, and those that do have not been able to incorporate recent Martian measurements \citep{jakosky_mars_1994}, or did not include isotopic constraints from $\ce{^{12}C}$, $\ce{^{13}C}$, and $\ce{^{40}Ar}$ \citep{kurokawa_lower_2018}. Like previous coupled models, the abundances and isotopic ratios of the atmospheric species are dynamically updated at each timestep according to a range of planetary processes. Their subsequent mixing ratios are then used to calculate the rate at which planetary processes impact each species in the atmosphere for the next timestep. Because we include multiple species at once, existing constraints on individual species are tested against each other, and new, comprehensive constraints on the Martian atmosphere emerge. Thus, solutions found with this model are the first to be self-consistent with respect to the evolution of the abundances of \ce{CO2}, \ce{N2}, and \ce{Ar}, and their constituent \ce{C}, \ce{N}, and \ce{Ar} isotopes in Mars's atmosphere. Other new aspects of our model include an obliquity and pressure dependent treatment of atmospheric collapse, and revised photochemical escape rates based on laboratory experiments and models. Like previous works \citep[e.g.,][]{kurokawa_lower_2018, hu_nitrogen-rich_2022}, we also present a comprehensive analysis of parameter space using a Markov Chain Monte Carlo method.

The rest of this paper is structured as follows. In Section \ref{sec:methods} we describe our model. In Section \ref{sec:results} we present constraints on the ancient Martian atmosphere and the results of our statistical analysis. In Section \ref{sec:disc} we discuss the implications of our model results for the ancient Martian atmosphere and compare our results to other studies. In Section \ref{sec:conc} we present conclusions from this study.

\section{Methods} \label{sec:methods}

\subsection{Model Overview} \label{subsec:model_overview}

\begin{figure*}[ht!]
\plotone{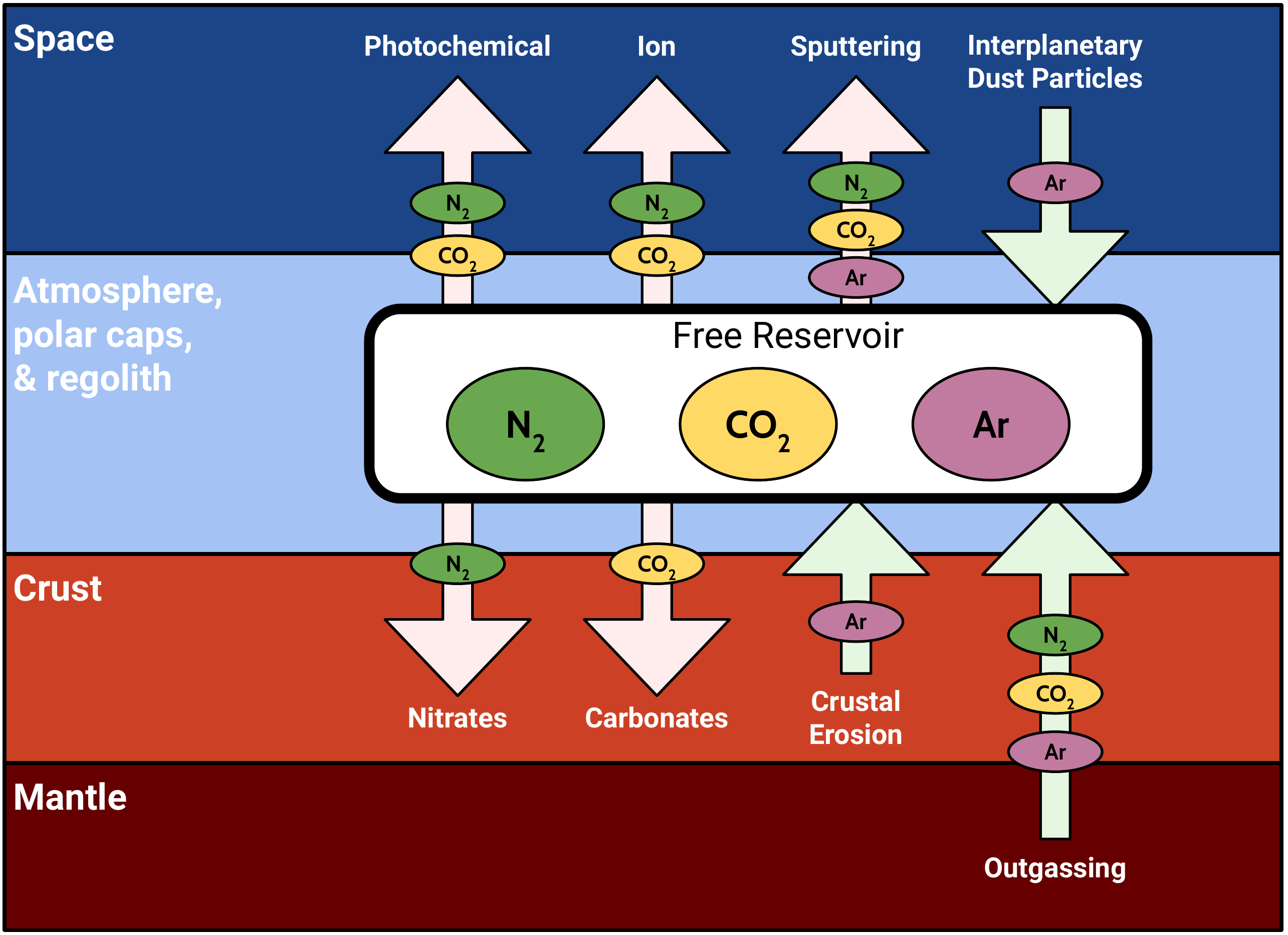}
\caption{Schematic diagram of the model developed in this work. Arrows pointing in and out of the free reservoir are sources and sinks of atmospheric species. This model tracks the abundance of each species and its isotopic composition in the free reservoir.}
\label{fig:box_mod}
\end{figure*}

The model we present tracks the abundance and isotopic composition of \ce{CO2}, \ce{N2}, and \ce{Ar} on Mars's surface from 3.8 Ga to present. 3.8 Ga is chosen as the model starting point because it is after the last major impact \citep[$\sim$3.9 Ga;][]{fassett_sequence_2011,robbins_large_2013}, and because geologic evidence indicates Mars's internal magnetic field should have ceased by this time \citep{lillis_rapid_2008}. This model is based on previous models presented in \citet{hu_tracing_2015} and \citet{hu_nitrogen-rich_2022}. The time evolution of the abundances of \ce{CO2}, \ce{N2}, and \ce{Ar} on Mars's surface are described by the following equations:

\begin{equation}
    \frac{d\ce{pCO2}}{dt} = F_{\rm og}^{\rm C} - F_{\rm sput}^{\rm C} - F_{\rm photo}^{\rm C} - F_{\rm ion}^{\rm C} - F_{\rm carb}^{\rm C}
    \label{eq:co2dif}
\end{equation}

\begin{equation}
    \frac{d\ce{pN2}}{dt} = F_{\rm og}^{\rm N} - F_{\rm sput}^{\rm N} - F_{\rm photo}^{\rm N} - F_{\rm ion}^{\rm N} - F_{\rm nit}^{\rm N}
    \label{eq:n2dif}
\end{equation}

\begin{equation}
    \frac{d\ce{pAr}}{dt} = F_{\rm og}^{\rm A} + F_{\rm ce}^{\rm A} + F_{\rm IDP}^{\rm A} - F_{\rm sput}^{\rm A} 
    \label{eq:Ardif}
\end{equation}

where $F^{i}_{j}$ represents the flux of species $i$ due to process $j$ in units mbar Myr$^{-1}$. $F_{\rm og}$ is volcanic outgassing, $F_{\rm sput}$ is escape caused by pickup-ion sputtering, $F_{\rm photo}$ is escape caused by photochemical reactions, $F_{\rm ion}$ is direct escape of ionized species, $F_{\rm carb}$ is the formation of carbonate minerals on the surface, $F_{\rm nit}$ is the formation of nitrate minerals on the surface, $F_{\rm ce}$ is release from crustal erosion, and $F_{\rm IDP}$ is the delivery of interplanetary dust particles (IDPs). \ce{pCO2}, \ce{pN2}, and \ce{pAr} are the partial pressures for each species in surface reservoirs that exchange on short timescales - referred to as the free reservoir (see Section \ref{subsec:freeres_collapse}). An overview of the model configuration is shown in Figure \ref{fig:box_mod}.

The evolution of the isotopic composition of \ce{CO2}, \ce{N2}, and \ce{Ar} is determined by the fluxes described above. Isotopic composition is calculated using delta notation reported in parts per thousand ($\permil$), defined for carbon in \ce{CO2} as:
\begin{equation}
    \delta^{13}\ce{C} = \frac{(^{13}\ce{C}/^{12}\ce{C})_{\rm Sample} - (^{13}\ce{C}/^{12}\ce{C})_{\rm Standard}}{(^{13}\ce{C}/^{12}\ce{C})_{\rm Standard}} \times 1,000
    \label{eq:C_delta_def}
\end{equation}
where the standard is VPDB with $(^{13}\ce{C}/^{12}\ce{C})_{\rm Standard} = (^{13}\ce{C}/^{12}\ce{C})_{\rm VPDB} = 0.0112372$ \citep{faure_principles_1991}. For nitrogen, $\delta^{15}\ce{N}$ is the enhancement of $^{15}\ce{N}/^{14}\ce{N}$ relative to the Earth's atmosphere, $(^{15}\ce{N}/^{14}\ce{N})_{\rm Standard} = (^{15}\ce{N}/^{14}\ce{N})_{\rm AIR} = 0.003676$ \citep{coplen_isotope-abundance_2002}. For argon, $\delta^{38}\ce{Ar}$ is the enhancement of $^{38}\ce{Ar}/^{36}\ce{Ar}$ relative to the solar value $(^{38}\ce{Ar}/^{36}\ce{Ar})_{\rm Standard} = (^{38}\ce{Ar}/^{36}\ce{Ar})_{\rm SUN} = 0.182$  \citep{vogel_argon_2011,pepin_helium_2012} and $\delta^{40}\ce{Ar}$ is the enhancement of $^{40}\ce{Ar}/^{36}\ce{Ar}$ relative to the modern Martian atmospheric value $(^{40}\ce{Ar}/^{36}\ce{Ar})_{\rm Standard} = (^{40}\ce{Ar}/^{36}\ce{Ar})_{\rm MARS} = 1900$ \citep{mahaffy_abundance_2013}. Sputtering, photochemical reactions, ion escape, and mineral formation are assumed to be Rayleigh fractionation processes with associated fractionation factors. Volcanic outgassing, crustal erosion, and the delivery of IDPs are mixing processes that introduce species of a distinct isotopic composition to the atmosphere, but do not have an inherent fractionation effect.

\subsection{The Free Reservoir and Atmospheric Collapse} \label{subsec:freeres_collapse}

The free reservoir includes all surface reservoirs that exchange particles on short timescales such that they are in isotopic and thermodynamic equilibrium over geologic time. \ce{N2} and \ce{Ar} do not condense into ice within the temperature and pressure range of Mars's history, so the free reservoir for these species only includes the atmosphere and any adsorption by the regolith. On the other hand, atmospheric \ce{CO2} can collapse to form polar ice caps in vapor equilibrium with the bulk atmosphere \citep{haberle_model_1994}. For this reason, the free reservoir of \ce{CO2} includes the atmosphere, the regolith, and the polar ice caps.

We consider the possibility that atmospheric collapse happened periodically throughout Mars's history. The criteria for atmospheric collapse in our model are based on 3D General Circulation Model (GCM) simulations done at the Laboratoire de M\'et\'eorologie Dynamique (the ``LMD'' model) \citep{forget_3d_2013}. For a given surface pressure, there is a critical value of obliquity: if Mars's obliquity is equal to or below this value, \ce{CO2} will collapse to the polar caps. According to this model, the atmosphere will not collapse regardless of obliquity when the surface pressure is between 600 mbar and 3 bar.

We must know if the obliquity at a given time is below the critical value in order to determine if the atmosphere was collapsed. The past obliquity of Mars is chaotic and not explicitly known prior to several million years ago. Furthermore, the timescale of Mars's obliquity cycle is much smaller than a typical model timestep, and thus the atmosphere may be both collapsed and uncollapsed in a single timestep \citep{ward_climatic_1974}. \citet{laskar_long_2004} put statistical constraints on the past obliquity of Mars and generate probability density functions for Mars's obliquity at discrete times as far back as 4 Ga. We integrate these functions to estimate the amount of time that Mars had an obliquity below the critical value during a timestep in our model and calculate the probability that the atmosphere was collapsed ($f_{\rm col}$):
\begin{equation}
    f_{\rm col}(t) = \int^{\phi_{\rm crit}}_0 {\rm P(\phi, t)} d\phi
    \label{eq:fcol_calc}
\end{equation}
where $t$ is the model time, $\phi$ is Mars's obliquity, $\phi_{\rm crit}$ is the critical obliquity from the GCM simulations, and ${\rm P(\phi, t)}$ is the probability density function from \citet{laskar_long_2004} that is closest to the model time. The probability of atmospheric collapse, $f_{\rm col}$, is interpreted as the fraction of each timestep spent in the collapsed state.

The flux of atmospheric escape for each species is calculated by taking the weighted average of the escape fluxes over a collapsed and an uncollapsed atmosphere. We first calculate the atmospheric escape fluxes for all species in the uncollapsed scenario, assuming the atmospheric \ce{pCO2} is equal to the \ce{CO2} contained in the free reservoir. We then calculate the atmospheric escape fluxes assuming the atmosphere is collapsed, where all \ce{CO2} in the free reservoir condenses onto the poles, and 6 mbar remains in the atmosphere to maintain vapor equilibrium. The fluxes calculated in these scenarios will be different because all atmospheric escape processes for each species depend on their mixing ratios, which depend on atmospheric \ce{pCO2}. To calculate the actual fluxes used to evolve the model, we take the weighted average of the fluxes with respect to the fraction of time spent in the collapsed state:
\begin{equation}
    F^{i} = f_{\rm col}F_{\rm collapsed}^{\rm i} + (1-f_{\rm col})F_{\rm uncollapsed}^{\rm i}
    \label{eq:flux_average}
\end{equation}
where $F^{i}$ is the flux of species $i$ from a source or sink. We refer to this as the pressure-obliquity dependent treatment of atmospheric collapse. To test model sensitivity, we also consider a scenario in which atmospheric collapse does not happen at all ($f_{\rm col} = 0$).

\subsection{Volcanic Outgassing} \label{subsec:outgassing}

Volcanic outgassing occurs when species dissolved in extrusive or intrusive Martian magma are exolved into the atmosphere. We model the flux of species $i$ outgassed into the free reservoir ($F_{\rm og}^{i}$) via the following equation \citep{hu_nitrogen-rich_2022}:
\begin{equation}
    F_{\rm og}^{i} =  V\rho_{\rm cr} X_{\rm mag}^{i} f_{\rm og},
\end{equation}
where $V$ is the crustal production rate, $\rho_{\rm cr}$ is the density of the crust (2900 kg m$^{-2}$), $X_{\rm mag}^{i}$ is the concentration of species {\it i} in the source magma, and $f_{\rm og}$ is a multiplication factor that accounts for the uncertainty in the crustal production rate, concentrations in the source magma, and the outgassing efficiency (including the extrusive-to-intrusive ratio). We employ a crustal production rate based on a combination of thermal evolution models and photogeological analysis \citep{greeley_magma_1991,grott_volcanic_2011} which is also adopted by a previous model of the argon isotope system \citep{slipski_argon_2016}. We consider the crater chronology of Mars's surface from \citet{hartmann_martian_2005} as the baseline case, and \citet{kallenbach_marsmoon_2001} as the variant \citep[see][Extended Data Figure 3b]{hu_nitrogen-rich_2022}.

The concentration of $\ce{CO2}$ in the source magma ($X_{\rm mag}^{\rm C}$) is solubility-limited and dependent on the redox state of the Martian mantle \citep{hirschmann_ventilation_2008, grott_volcanic_2011}. These interior models indicate that $X_{\rm mag}^{\rm C}$ ranges from 5 ppm at mantle oxygen fugacity of IW-1 to as much as 1,000 ppm at IW+1, so we explore this range. We assume that the concentration of $\ce{N2}$ in the source magma ($X_{\rm mag}^{\rm N}$) is not solubility-limited, and we adopt the ``Silicate Earth" concentration, $X_{\rm mag}^{\rm N} = 1.9$ ppm \citep{marty_nitrogen_2003}. The concentration and isotopic composition of \ce{Ar} in the source magma is discussed in Section \ref{subsec:argon_system}.

The introduction of species $i$ into the free reservoir via volcanic outgassing has an impact on isotopic composition due to bulk mixing. This is described by the equation:

\begin{equation}
    \delta^{x}i = \frac{P_{\rm og}\delta^{x}i_{\rm mag} + P_{\rm free}\delta^{x}i_{\rm free}}{P_{\rm og}+P_{\rm free}},
\end{equation}

where $\delta^{x}i$ is the updated isotopic composition of species \textit{i} in the free reservoir, $P_{\rm og}$ is the partial pressure of species \textit{i} outgassed per time step, $P_{\rm free}$ is the partial pressure of species \textit{i} in the free reservoir before the outgassing occurs, $\delta^{x}i_{\rm mag}$ is the isotopic composition of species \textit{i} in the source magma, and $\delta^{x}i_{\rm free}$ is the isotopic composition of species \textit{i} in the free reservoir before the outgassing occurs. We assume $\delta^{13}\ce{C}_{\rm mag} =$ -25$\permil$ based on analysis of the mantle degassed $\ce{CO2}$ from the magmatic component of the SNC (shergottites, nakhlites, chassignites) meteorites \citep{wright_chassigny_1992}. We assume $\delta^{15}\ce{N}_{\rm mag} =$ -30$\permil$ based on measurements of the Martian meteorite ALH 84001 \citep{mathew_early_2001}. 

\subsection{The Argon Isotope System and Crustal Erosion} \label{subsec:argon_system}

The three most abundant argon isotopes on present-day Mars are \ce{^{36}Ar}, \ce{^{38}Ar}, and \ce{^{40}Ar} \citep{atreya_primordial_2013,mahaffy_abundance_2013}. Thus, argon on the surface of Mars can be characterized by a total partial pressure of the 3 isotopes combined, and two values that relate to isotopic composition - $\delta^{38}\ce{Ar}$ and $\delta^{40}\ce{Ar}$, both of which track enrichment relative to \ce{^{36}Ar}. \ce{^{36}Ar} and \ce{^{38}Ar} are both stable isotopes and \ce{^{40}Ar} is radiogenically produced from the decay of \ce{^{40}K}. The stable Ar isotopes are treated similarly to \ce{CO2} and \ce{N2}. Following \citet{slipski_argon_2016}, we derive Martian source magma abundances of $\ce{^{36}Ar}$ and $\ce{^{38}Ar}$ from Earth's atmosphere. Thus, we assume $X_{\rm mag}^{\rm A36} = 3.46 \times 10^{-5}$ ppm \citep{marty_origins_2012} and $\ce{^{36}Ar}/\ce{^{38}Ar}$ = 5.305 \citep{lee_redetermination_2006}, which is equivalent to assuming $\delta^{38}\ce{Ar}_{\rm mag}$ = 36$\permil$ relative to the solar abundance.

The treatment of $\ce{^{40}Ar}$ is more complicated because it is sourced from the radioactive decay of \ce{^{40}K}. The radiogenic production of $\ce{^{40}Ar}$ occurs almost entirely in the crust and mantle because potassium is highly refractory. Thus, the concentration of \ce{^{40}Ar} in the source magma for volcanic outgassing is directly related to the magma concentration of \ce{^{40}K} and changes as the planet evolves. Additionally, \ce{^{40}K} is sequestered into the Martian crust during volcanic emplacement, and will then decay into \ce{^{40}Ar} in the crust. This crustal \ce{^{40}Ar} can then be released by crustal erosion and represents another source of \ce{^{40}Ar} to the atmosphere. Once in the atmosphere, \ce{^{40}Ar} is treated the same as the other argon isotopes.

We follow the model of \citet{slipski_argon_2016} to calculate the evolution of \ce{^{40}Ar}. Their model tracks the abundance of \ce{^{40}K} and \ce{^{40}Ar} in the crust and mantle, and includes decay processes, volcanic emplacement, and crustal erosion. Following \citet{slipski_argon_2016}, we assume a mantle \ce{^{40}K} concentration of 0.4 ppm at 4.4 Ga, which decreases with time as it decays and is sequestered into the crust. The crustal production rate and volcanic outgassing multiplier in our model determines how much \ce{^{40}K} is lost from the mantle to the crust. We extend our model crustal production rate from 3.8 Ga to 4.4 Ga by employing the interior model of \citet{grott_volcanic_2011}. We do not need to extend any other model processes to the time before our model domain because the \ce{^{40}K} and \ce{^{40}Ar} interior evolution does not depend on any other aspects of the modeled system. In the end, we obtain the \ce{^{40}Ar} source magma concentration ($X_{\rm mag}^{\rm A40}$) as a function of time from 3.8 Ga to present, which we use to calculate the \ce{^{40}Ar} volcanic outgassing flux.

During volcanic emplacement, \ce{^{40}K} is sequestered into the crust, where it eventually decays into \ce{^{40}Ar}. Following \citet{leblanc_mars_2012} and \citet{slipski_argon_2016}, we assume \ce{^{40}K} is enriched in the crust by a factor of 5 because it is an incompatible element and will concentrate in the melt during volcanic emplacement. \citet{slipski_argon_2016} show that the release of crustal \ce{^{40}Ar} to the atmosphere (termed crustal erosion) is critical for reproducing the modern atmospheric composition, but the process through which the release occurs is poorly understood. Multiple methods of crustal erosion have been hypothesized to occur on Earth, some of which depend on groundwater and some of which don't \citep{watson_40ar_2007}, but none have been confirmed. For simplicity, we model the crustal erosion of \ce{^{40}Ar} as a constant rate throughout Mars's history. Uncertainty in this approach is absorbed into the sputtering multiplier and the crustal erosion multiplier, which is described below. To calculate the rate of \ce{^{40}Ar} supplied to the atmosphere from crustal erosion, we first calculate the total amount of \ce{^{40}Ar} produced in the crust over Mars's history in the same way we calculate the evolution of the mantle \ce{^{40}Ar} concentration. We then calculate the amount of \ce{^{40}Ar} released to the atmosphere from crustal erosion as some percentage of the total produced crustal \ce{^{40}Ar}. $f_{\rm ce}$ is the parameter that determines this percentage, and varies from 0 to 1. The amount of \ce{^{40}Ar} released from crustal erosion is then spread over the model time domain as a constant source flux. 

\subsection{Mass-Dependent Separation above the Homopause} \label{subsec:mass_dep_sep}

Sputtering, photochemical reactions, and direct loss of ionized species all take place at altitudes above the homopause, where the atmosphere is no longer well-mixed and each species takes on its mass-dependent scale height. This has implications for fractionation due to atmospheric escape processes because the lighter version of a given species will accumulate higher in the atmosphere, where it is more likely to be ejected. This is a diffusive fractionation effect, which is different from the fractionation effect inherent to a given loss process. The total fractionation factor for an atmospheric escape process is the product of the diffusive and the inherent fractionation factors. The fractionation factor due to mass-dependent separation above the homopause ($\alpha^{i,j}_{\rm sep}$) is: 
\begin{equation}
    \alpha^{i,j}_{\rm sep} = \exp\bigg(\frac{-g\Delta m_{i,j}\Delta z}{kT}\bigg), 
    \label{eq:diffu}
\end{equation}
where $g$ is Mars's surface gravity, $\Delta m_{i,j}$ is the mass difference of the particles specified by $i$ and $j$, $\Delta z$ is the distance from the homopause to the altitude of escaping particles (i.e. the exobase), $k$ is the Boltzmann constant, and $T$ is the mean temperature of the thermosphere. 

The quantity $\Delta z/T$ is important for determining the fractionation factor due to mass-dependent separation above the homopause. This parameter has been recently constrained by MAVEN measurements of argon in Mars's upper atmosphere \citep{jakosky_mars_2017}. Analysis of the MAVEN data shows that $\Delta z/T$ varies on short timescales in the range 0.2-0.5 km K$^{-1}$. Thus, we take this quantity as a free parameter within this range that is constant over our model time domain. By employing the same value of $\Delta z/T$ for carbon, nitrogen, and argon, we implicitly assume that all atmospheric sinks are generating escaping particles at the same source altitude. This assumption is justified by recent investigations. The MAVEN measurements of argon indicate that the altitude of the exobase is in the range 140 - 200 km \citep{jakosky_mars_2017}. This range is broadly consistent with models of photochemically escaping carbon \citep{lo_carbon_2021} and nitrogen \citep{fox_production_1993}, and observations of ion escape \citep{jakosky_maven_2015}.

\subsection{Sputtering} \label{subsec:sputtering}

Atmospheric loss via pickup ion sputtering occurs when oxygen ions are accelerated by the solar wind and collide with other particles in the upper atmosphere, ejecting them from Mars's gravity \citep{kass_change_1999, leblanc_role_2002}. This process occurs because Mars does not have an intrinsic magnetic field to protect the atmosphere from the solar wind during the modeled period \citep{lillis_rapid_2008}. This process was likely more effective at early epochs when the solar extreme ultraviolet (EUV) flux was higher. To model the sputtering flux ($F_{\rm sp}^{i}$), we employ the following equation:
\begin{equation}
    F_{\rm sp}^{i}= F_{\rm sp}^{\rm C} \frac{Y^{i}}{Y^{\rm C}} \frac{X^{i}_{\rm atm}}{X^{\rm C}_{\rm atm}} \alpha^{i, \rm C}_{\rm sep} \frac{1}{\alpha_{\rm dil}}f_{\rm sp},
    \label{eq:sput}
\end{equation} 
where $F_{\rm sp}^{\rm C}$ is the sputtering rate of \ce{CO2}, $Y_i$ is the yield for species \textit{i}, $X^{i}_{\rm atm}$ is the concentration of species \textit{i} in the bulk atmosphere, $\alpha^{i, \rm C}_{\rm sep}$ is the separation factor between species \textit{i} and \ce{CO2} at the exobase (Eq. \ref{eq:diffu}), $\alpha_{\rm dil}$ is a factor that accounts for the dilution of species \textit{i} by other species at the exobase (Eq. \ref{eq:dil}), and $f_{\rm sp}$ is a multiplication factor to account for uncertainty in the parameterization. There is no inherent isotopic fractionation due to sputtering because of the high energy imparted on the escaping particles. All fractionation from sputtering is due to mass-dependent separation above the homopause.

To calculate $F_{\rm sp}^{\rm C}$, we employ 3D Monte Carlo simulations \citep{leblanc_role_2002} fitted to the functional form:
\begin{eqnarray} 
    F_{\rm sp}^{\rm C} = \exp(-0.462\ln(F_{\rm EUV}/F_{\rm 0,EUV})^2 + \nonumber \\ 5.086\ln(F_{\rm EUV}/F_{\rm 0,EUV}) + 53.49)
\end{eqnarray}
in units of particles per second, where $F_{\rm EUV}$ is the solar EUV flux and subscript 0 indicates the present-day value. We model the evolution of the EUV flux as $F_{\rm EUV} \propto t^{-1.23 \pm 0.1}$, where $t$ is the age, and the EUV flux is larger at earlier times \citep{ribas_evolution_2005,claire_evolution_2012, tu_extreme_2015}. We scale the above \ce{CO2} sputtering rate to \ce{N2} and \ce{Ar} using the ratio of the yields calculated in Monte Carlo simulations by \citet{jakosky_mars_1994}: $Y^{C} = 0.7$, $Y^{N} = 2.4$, and $Y^{A} = 1.4$.

The sputtering rate for a given species is reduced by the presence of other species that can collide with the incident ions. This is captured by the dilution factor ($\alpha_{\rm dil}$):
\begin{equation}
    \alpha_{\rm dil} \equiv 1 + \sum_i \frac{X^i_{\rm atm}}{X^{\rm C}_{\rm atm}} \alpha^{i, \rm C}_{\rm sep},
    \label{eq:dil}
\end{equation}
where the sum is over the relevant species in Mars's atmosphere and $\alpha^{i, \rm C}_{\rm sep}$ is the diffusion separation between species $i$ and \ce{CO2} (Eq. \ref{eq:diffu}). In addition to \ce{CO2}, \ce{N2}, and \ce{Ar}, we include the minor species in Mars's atmosphere in this sum: \ce{O} and \ce{CO}, which we assume to have abundances of 0.16\%, and 0.06\% by volume, respectively. 

\subsection{Photochemical Escape} \label{subsec:photochem}

\subsubsection{Carbon} \label{subsubsec:photo_c}

We scale the photochemical escape rate of carbon ($F^{\rm C}_{\rm photo}$) by the evolution of the solar lyman continuum flux and the atmospheric abundance of \ce{CO2}:
\begin{equation} 
    F_{\rm photo}^{\rm C} = F_{\rm 0, photo}^{\rm C} \left( \frac{F_{\rm LYM}}{F_{\rm 0, LYM}} \right)^{a_{\rm LYM}} \frac{X^{\ce{C}}_{\rm atm}}{X^{\ce{C}}_{\rm 0, atm}} f_{pC}
\end{equation}
where $F_{\rm LYM}$ is the solar Lyman continuum flux, $X^{\ce{CO2}}_{\rm atm}$ is the mixing ratio of \ce{CO2} in the bulk atmosphere, $a_{\rm LYM}$ is a power-law index, $f_{pC}$ is a multiplication factor, and subscript $0$ indicates the present-day value. The power-law index and the multiplication factor are included to capture the uncertainties in the total escape rate as well as its dependence on the evolution of the solar lyman continuum flux. We adopt $F_{\rm LYM} \propto t^{-0.86 \pm 0.1}$ based on observations of young solar-like stars and the wavelength ranges that drive the majority of photochemical carbon escape \citep{ribas_evolution_2005, claire_evolution_2012, lo_carbon_2021}.

We break down the total photochemical escape rate of carbon into the five most efficient escape reactions according to recent 1D Monte Carlo photochemical models \citep{lo_carbon_2021}:
\begin{equation}
    F_{\rm photo}^{\rm C} = F_{\rm pd, cd}^{\rm C} + F_{\rm pd, cm}^{\rm C} + F_{\rm el, cd}^{\rm C} + F_{\rm dr, cm}^{\rm C} + F_{\rm pi, cm}^{\rm C}
\end{equation}
where the terms on the right hand side correspond to rates of escape from photodissociation of \ce{CO2} (44.5$\%$ of total escaping C at modern day), photodissociation of \ce{CO} (20$\%$), electron impact of \ce{CO2} (12.3$\%$), dissociative recombination of \ce{CO+} (13.1$\%$), and photoionization of \ce{CO} (8.7$\%$), respectively. Escape of carbon via photodissociation of \ce{CO2} was previously unconsidered on Mars, but the recent results from the photoechemical model of \citet{lo_carbon_2021} suggest that it is actually the dominant escape mechanism. These models calculate the total rate of escaping carbon to be $F_{\rm 0, photo}^{\rm C} = 4.68 \times 10^{23}$ (C atoms s$^{-1}$), and the five reactions considered here are responsible for 98.6\% of the escaping carbon. We obtain the escape fluxes used here by averaging over the extremes of Mars's orbit and the solar cycle, using the larger \ce{C}-\ce{CO2} collisional cross section  \citep[][their Table 2]{lo_carbon_2021}. 

To calculate the overall inherent fractionation factor due to photochemical loss of carbon, we take the weighted average of the fractionation factors of the individual reactions. We calculate the fractionation factor of \ce{CO2} photodissociation to be 0.68 (see Section \ref{subsec:photodiss_c}). The fractionation factors of \ce{CO} photodissociation and \ce{CO+} dissociative recombination have been previously calculated to be 0.6 and 0.8, respectively \citep{ fox_15_1997,hu_tracing_2015}. Because it has not been calculated, we approximate the fractionation factor of electron impact of \ce{CO2} by assuming it is equivalent to the fractionation factor of \ce{CO+} dissociative recombination (0.8). We assume the fractionation factor of \ce{CO} photoionization is 1 because photoionization photons are typically much more energetic than the escape threshold. The weighted average fractionation factor for photochemical loss of carbon is thus 0.73.

\subsubsection{Fractionation due to Photodissociation of \ce{CO2}} \label{subsec:photodiss_c}

\begin{figure}
\plotone{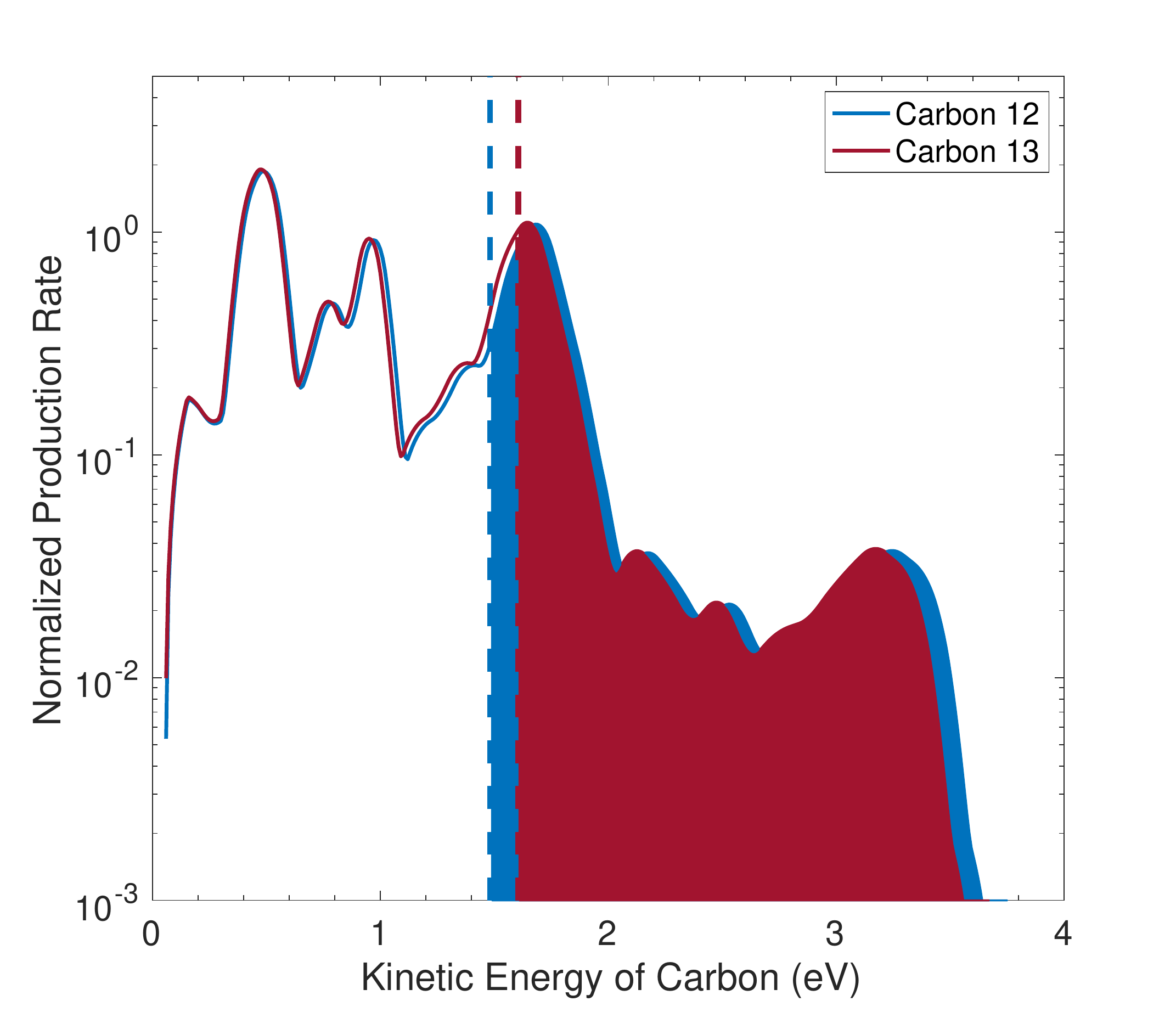}
\caption{Energy distribution of carbon atoms produced by the photodissociation of \ce{CO2} in Mars's upper atmosphere. The required energies to escape Mars's gravity for the carbon isotopes are shown by the dashed lines. The shaded regions represent the portion of carbon isotopes that are ejected from Mars's atmosphere.}
\label{fig:pd_frac}
\end{figure}

Because \ce{CO2} photodissociation is now predicted to be the dominant loss mechanism of carbon on Mars \citep{lo_carbon_2021}, we present the first calculation of the \ce{CO2} photodissociation fractionation factor. We utilize the Photochemical Isotope Effect (PIE) method \citep{hu_tracing_2015, hu_nitrogen-rich_2022} to appropriately distribute energy from the incident photons into the dissociating \ce{CO2} molecule. Energy from the incident photons that is above the threshold energy of the dissociation channels is partitioned into kinetic energy of the carbon and oxygen atoms from the dissociated \ce{CO2} molecule. To determine the production rate of carbon atoms as a function of nascent kinetic energy, we employ a globally-averaged 1D coupled ion-neutral photochemical model that spans from the surface of Mars to 240 km, at 1 km resolution \citep{lo_carbon_2020,lo_carbon_2021}. This model takes as input the solar spectrum, the extremes of the solar cycle and Mars's orbital distance, reaction rates among atmospheric species, cross sections for photochemical interaction, and branching ratios of dissociation pathways.

The results of this calculation are shown in Figure \ref{fig:pd_frac}. \ce{^{12}C} is preferentially ejected from the atmosphere relative to \ce{^{13}C} because it has a lower escape energy and it obtains a higher portion of excess kinetic energy following photodissociation. We obtain the fractionation factor by integrating the shaded regions and dividing the fraction of escaping \ce{^{13}C} by the fraction of escaping \ce{^{12}C}. Via this method, we calculate the fractionation factor of escaping carbon due to photodissociation of \ce{CO2} to be 0.68.

\subsubsection{Nitrogen} \label{subsec:photo_n}

We scale the photochemical escape rate of nitrogen ($F^{\rm N}_{\rm photo}$) by the evolution of the solar EUV flux and the atmospheric abundance of \ce{N2}:
\begin{equation} 
    F_{\rm photo}^{\rm N} = F_{\rm 0, photo}^{\rm N} \left( \frac{F_{\rm EUV}}{F_{\rm 0, EUV}} \right)^{a_{\rm EUV}} \frac{X^{\ce{N}}_{\rm atm}}{X^{\ce{N}}_{\rm 0, atm}} f_{pN}
    \label{eq:nphoto}
\end{equation}
where $X_{\ce{N2}}$ is the mixing ratio of \ce{N2} in the bulk atmosphere, $a_{\rm EUV}$ is a power-law index, $f_{pN}$ is a multiplication factor, and subscript 0 indicates the present-day value. The power-law index and the multiplication factor are included to capture the uncertainties in the total escape rate as well as its dependence on the evolution of the solar EUV flux. The evolution of the EUV flux is modeled as $F_{\rm EUV} \propto t^{-1.23 \pm 0.1}$ (See Section \ref{subsec:sputtering}). \citet{tu_extreme_2015} suggest that this exponent can actually vary in the range -0.96 to -2.15 depending on the initial solar rotation rate. By compounding the assumed fixed value of -1.23 with the $a_{\rm EUV}$ exponent, we explore a wide parameter range that covers most of the possible solar EUV evolution scenarios. This approach is justified because previous models that explore the full range of EUV evolution scenarios show that the atmospheric escape of nitrogen is insensitive to this effect \citep{hu_nitrogen-rich_2022}.

We break down the photochemical escape rate into different mechanisms because they each have different fractionation factors:
\begin{equation}
    F_{\rm photo}^{\rm N} = F_{\rm pd}^{\rm N} + F_{\rm dr}^{\rm N} + F_{\rm other}^{\rm N}
\end{equation}
where $F_{\rm pd}^{\rm N}, F_{\rm dr}^{\rm N}$, and $F_{\rm other}^{\rm N}$ are the rates of escape from photodissociation and photoionization, dissociative recombination, and other chemical reactions, respectively. The present-day escape rates of dissociative recombination and other chemical reactions are baselined in \citet{fox_production_1993}, and the present-day escape rate of photodissociation is calculated in \citet{hu_nitrogen-rich_2022}. We assume the scaling in Eq. \ref{eq:nphoto} applies equally to these processes. 

The inherent fractionation factors of photodissociation \citep{hu_nitrogen-rich_2022} and dissociative recombination \citep{fox_15_1997} are previously calculated ($\alpha_{\rm pd}^{\rm N} = 0.29$ and $\alpha_{\rm dr}^{\rm N} = 0.58)$. The inherent fractionation factor due to other chemical reactions is assumed to be unity ($\alpha_{\rm dr}^{\rm N} = 1$) because the dominant reactions supply much more kinetic energy than is required for escape \citep{fox_nitrogen_1983}.

\subsection{Ion Escape} \label{subsec:ion}

Mars is susceptible to atmospheric erosion via ion escape, where charged atmospheric particles in the upper atmosphere are ejected via direct interaction with the solar wind. This process has been investigated directly by Mars Express and the Mars Atmosphere and Volatile Evolution Orbiter (MAVEN) \citep{barabash_martian_2007, jakosky_maven_2015}, and the measured escape rates of \ce{CO2+}, \ce{O2+}, and \ce{O+} generally agree with magneto-hydrodynamic (MHD) model predictions \citep{ma_ion_2007}. We extrapolate the escape rates back in time by fitting the present day measurement-validated MHD escape rates to a power law of the solar age via a parametric model \citep{manning_parametric_2011}. The ion loss rate of carbon, $F_{\rm ion}^{\rm C}$, is modeled as: 
\begin{equation}
    F_{\rm ion}^{\rm C} =  \frac{X^{\ce{C}}_{\rm atm}}{X^{\ce{C}}_{\rm 0, atm}} F_{\rm 0,ion}\big(\ce{CO2+}\big) \bigg( \frac{t}{4500} \bigg)^{-3.51},\label{eq:ioncarbon}
\end{equation}
where $F_{\rm 0,ion}\big(\ce{CO2+}\big)$ is the present-day ion escape rate of \ce{CO2+} from the parametric model and the last term accounts for the evolution of the solar EUV flux with the power-law index from the parametric model. The ion loss rate of nitrogen, $F_{\rm ion}^{\rm N}$, is proportional to the ion loss rate of carbon. It is modeled as: 
\begin{equation}
    F_{\rm ion}^{\rm N} = \frac{X^{\ce{N}}_{\rm atm}}{X^{\ce{N}}_{\rm 0, atm}} \frac{X^{\ce{N2+}}_{0, \rm atm}}{X^{\ce{CO2+}}_{0, \rm atm}} F_{\rm ion}^{\rm C},\label{eq:ionnitrogen}
\end{equation}
where $X^{\ce{N2+}}_{0, \rm atm}$ and $X^{\ce{CO2+}}_{0, \rm atm}$ are the present-day mixing ratios measured by MAVEN at the altitude of 160 km \citep{bougher_early_2015}. There is no inherent fractionation due to ion escape because it is much more energetic than the required escape energy for carbon and nitrogen. All fractionation from ion loss is due to mass-dependent separation above the homopause. We do not include an ion loss multiplier because the rate of escaping particles due to ion loss is several orders of magnitude lower than the other atmospheric escape processes. Thus, uncertainty in the ion escape rate can be incorporated into uncertainty in the other atmospheric escape processes.

\subsection{Mineral Deposition}

\subsubsection{Carbonates}

Following \citet{hu_tracing_2015}, we assume 300-1400 mbar \ce{CO2} is sequestered globally as carbonates on Mars. Carbonate minerals have been detected by orbital remote sensing of the Martian surface and \textit{in situ} measurements \citep{bandfield_spectroscopic_2003, ehlmann_orbital_2008, morris_identification_2010, niles_geochemistry_2013}.  Although they are detected, global infrared remote sensing observations indicate that carbonates are scarce on the surface compared to other secondary minerals \citep{ehlmann_mineralogy_2014}. The limits we apply to the global mass of carbonates are consistent with this. The upper limit of 1400 mbar assumes carbonates are 5 wt\% in the crust to a depth of 500 m, which is the maximum amount that would not be detectable from remote sensing. The lower limit of 300 mbar assumes carbonates are 1 wt\% in the crust to the same depth, which is a more plausible value that is consistent with observation \citep{niles_geochemistry_2013}. There is also evidence for carbonates sequestered into the deep crust, although the exact size and method of deposition is uncertain \citep{wray_orbital_2016}. We do not explicitly consider this reservoir, however its potential presence should be taken into account when interpreting our results. We explore this range of values with the parameter C-DEP, the equivalent atmospheric \ce{CO2} pressure sequestered in carbonates deposited before 3 Ga.

We model the rate of carbonate deposition as a step function, comprised of an early carbonate formation rate, a late carbonate formation rate, and a time of transition. We employ this method because we do not explicitly calculate surface temperature, and thus cannot accurately calculate the rate of deposition according to aqueous chemical kinetics. Additionally, taking the step function approach minimizes the number of free parameters and allows straightforward comparison to the geologic record. In reality, there would be short-timescale variability in the deposition rate depending on the transient presence of water.

We consider two early environments in which carbonate deposition occurred: open-water systems (OWS) and shallow sub-surface aquifers (SSA). The baseline scenario is deposition in open water systems (e.g. lakes and ponds), where the carbonate formed is $\sim$10 $\permil$ enriched in the heavy isotope compared to the atmosphere \citep{faure_principles_1991} and the time of transition from high to low formation rate is 3.5 Ga i.e., approximately between the Noachian and Hesperian. The variant scenario is deposition in shallow sub-surface aquifers, where the carbonate formed is up to $\sim$60 $\permil$ enriched compared to the atmosphere \citep{halevy_carbonates_2011} and the time of transition is 3 Ga i.e., by the end of the Hesperian. Following the transition to low carbonate formation, we assume 7 mbar of \ce{CO2} is lost to carbonate formation throughout the rest of Mars's history \citep{hu_tracing_2015}.

\subsubsection{Nitrates}

Evolved gas experiments with Sample Analysis at Mars (SAM) on the Curiosity rover have measured nitrate in Mars's soil and rocks \citep{stern_evidence_2015, sutter_evolved_2017}. Thus, we include it in the model as a sink for atmospheric \ce{N2}. Like carbonate deposition, we model the rate of nitrate deposition as a step function, comprised of an early nitrate formation rate, a late nitrate formation rate, and a time of transition. We assume there is no fractionation of the free nitrogen reservoir due to the formation and deposition of nitrates \citep{hu_nitrogen-rich_2022}.

The time of transition in the model step function is fixed at the Hesperian-Amazonian boundary, 3.0 Ga. In the Amazonian period, we estimate the amount of \ce{N2} deposited as nitrates by adopting the \ce{NO3} concentration in the Rocknest samples \citep{sutter_evolved_2017} and assuming a globally average regolith depth of 10 meters. This corresponds to $0.03 \pm 0.01$ mbar of \ce{N2}, which is then spread evenly over the Amazonian. The nitrate concentration in Noachian and Hesperian aged rocks is less constrained. Following \citet{hu_nitrogen-rich_2022}, we use a default rock concentration of 300 ppm by weight in \ce{NO3}, and assume an equivalent depth ($d$). The equivalent depth can be as large as 1000 m, but due to the relative insensitivity of the atmospheric evolution to nitrate deposition \citep{hu_nitrogen-rich_2022} we fix this value at $d = 50$ m. The removal rate is then calculated by evenly distributing the nitrate deposited over the Noachian and Hesperian periods (3.0-3.8 Ga) to derive a constant rate.

\subsection{Interplanetary Dust Particles (IDPs)}

It has been suggested that the accretion of interplanetary dust particles (IDPs) is responsible for a non-negligable amount of noble gas addition to the Martian atmosphere over its history \citep{flynn_contribution_1997, kurokawa_lower_2018}. The flux of accreting dust particles on Mars has never been directly measured, so we employ the model of \citet{flynn_contribution_1997}. This model utilizes measurements of IDPs in Earth’s stratosphere to calculate the IDP accretion rate, and it is consistent with IDP measurements made by Pioneer 10 and 11 in interplanetary space \citep{humes_results_1980}. The combined rate of accretion for $\ce{^{36}Ar}$ and $\ce{^{38}Ar}$ on Mars is calculated to be $1.12 \times 10^3$ g yr$^{-1}$. The $\ce{^{36}Ar}/\ce{^{38}Ar}$ ratio is not well constrained from this experiment, so we assume the solar value of 5.5. We assume there is negligible \ce{^{40}Ar} in accreted IDPs. The rate of Ar addition via IDP accretion is assumed to be a constant in our model's time domain. 

\subsection{Impactors}

Asteroids and comets (i.e., impactors) may have been important for the delivery and removal of volatiles in the Martian atmosphere in its early history \citep{svetsov_atmospheric_2007, de_niem_atmospheric_2012, slipski_argon_2016, kurokawa_lower_2018}. Because our model starts at 3.8 Ga, it does not include the proposed large impactor flux of the Late Heavy Bombardment (LHB). The uncertainty in the effect of the LHB is then absorbed into the partial pressure of the species at 3.8 Ga in our model.

We assume that the delivery and erosion of volatiles via asteroid impacts is negligible during our model time domain. Erosion of \ce{CO2} via impactors is potentially important during and before the LHB, but not after 3.8 Ga \citep{jakosky_co2_2019}. Recent atmospheric evolution models suggest that the delivery and erosion of \ce{N2} is negligible in comparison to other sources and sinks after 3.8 Ga \citep{kurokawa_lower_2018}. \citet{slipski_argon_2016} estimate an upper limit of argon delivery by directly analyzing post-LHB Martian cratering records \citep{robbins_new_2012, tanaka_digital_2014}. They find that 1$\%$ of the present-day $\ce{^{36}Ar}$ could be delivered via asteroids in the Amazonian, and 2.5$\%$ in the Hesperian. When spread over the modeled period, the rate of argon addition is negligible in comparison to other model uncertainties. For these reasons, we ignore asteroids and comets as a source or sink of atmospheric volatiles in our model.

\subsection{Present-day Mars and 3.8 Ga Conditions} \label{subsec:init_conds}

\begin{deluxetable*}{cccc}[ht!]
\tablenum{1}
\label{table:mcmc_params}
\tablecaption{Parameter ranges explored by the baseline and extended range MCMC searches.}
\tablehead{\colhead{Free Parameter} & \colhead{Description} & \colhead{Baseline MCMC} & \colhead{Extended Range MCMC}}
\startdata
$f_{og}$ & Volcanic Outgassing Multiplier &  0.5-2 &  0.01-20 \\
$f_{sp}$ & Sputtering Multiplier & 0.5-2 & 0.01-20 \\
$f_{pC}$ & Photochemical Multiplier - Carbon & 0.5-2 & 0.01-20 \\
$f_{pN}$ & Photochemical Multiplier - Nitrogen & 0.5-2 & 0.01-20 \\
$f_{ce}$ & Crutal Erosion Multiplier & 0-1 & 0-1 \\
$\Delta$z/T & Upper atmosphere structure (km K$^{-1}$) & 0.2-0.5 & 0.2-0.5 \\
$X^{\rm C}_{\rm mag}$ & \ce{CO2} concentration in source magma (ppm) & 5-1000 & 5-1000 \\
$a_{\rm EUV}$ & EUV power law index & 0.5-3 & 0.5-3 \\
$a_{\rm LYM}$ & Lyman continuum power law index & 0.5-3 & 0.5-3 \\
\enddata
\end{deluxetable*}

\begin{deluxetable*}{r|cc}[ht!]
\tablenum{2}
\label{table:variants}
\tablecaption{Model variants tested in the MCMC search. Each variant is tested one by one, and all other aspects of the model configuration follow the baseline scenario. See Section \ref{sec:methods} for more details.}
\tablehead{\colhead{} & \colhead{Baseline MCMC} & \colhead{Variant MCMC}}
\startdata
Volcanism Profile & \citet{hartmann_martian_2005} & \citet{kallenbach_marsmoon_2001} \\
Atmospheric Collapse  & Pressure and Obliquity Dependent & None \\
Carbonate Deposition Scenario & Open Water System & Shallow Subsurface Aquifer \\
Isotopic Composition at 3.8 Ga & $\delta^{15}\ce{N} = -30\permil$, $\delta ^{38}\ce{Ar} = 36\permil$ & $\delta^{15}\ce{N} = 300\permil$, $\delta ^{38}\ce{Ar} = 1255\permil$ \\
\enddata
\end{deluxetable*}

To compare the model results to present-day Martian conditions, we must determine the size and isotopic composition of the modern free reservoir for carbon, nitrogen, and argon. The atmospheric abundances of carbon, nitrogen, and argon have been directly measured on Mars \citep{franz_reevaluated_2015}: 6 mbar \ce{CO2}, 0.12 mbar \ce{N2}, 0.12 mbar \ce{Ar}. We assume the present day polar cap reservoir contains 7 mbar \ce{CO2} \citep{smith_time_2009,phillips_massive_2011}. The size of the regolith reservoir for carbon, nitrogen, and argon is uncertain because we do not know the depth to which it extends or the grain size, which determines the available surface area for adsorption of a given species. We consider a range of possible scenarios assuming a maximum regolith depth of 100 meters and a grain surface area between 20 and 100 m$^2$ g$^{-1}$ \citep{zent_fractionation_1994, zent_simultaneous_1995}. Summing these reservoirs, we find the present-day total inventories are as follows: $\ce{pCO2} = 34 \pm 20$ mbar, $\ce{pN2} = 0.25 \pm 0.13$ mbar, and $\ce{pAr} = 0.26 \pm 0.15$ mbar. The uncertainty represents the allowed range of model values that would constitute a solution and can be considered $3\sigma$.

We assume the present-day isotopic composition of each species is equal to the directly measured isotopic composition of Mars's atmosphere \citep{mahaffy_abundance_2013, atreya_primordial_2013} and is uniform within the free reservoir ($1\sigma$): $\delta ^{13}\ce{C} = 46 \pm 2 \permil$, $\delta ^{15}\ce{N} = 572 \pm 82 \permil$, $^{36}\ce{Ar}/^{38}\ce{Ar} = 4.2 \pm 0.1$ ($\delta ^{38}\ce{Ar} = 310 \pm 30 \permil$), and  $^{40}\ce{Ar}/^{36}\ce{Ar} = 1900 \pm 300$ ($\delta ^{40}\ce{Ar} = 0 \pm 79 \permil$). The isotopic composition of the three species at 3.8 Ga is required to evaluate the model evolution. In the baseline scenario, we assume $\delta^{13}\ce{C} = -25\permil, \delta^{15}\ce{N} = -30\permil,$ and $\delta ^{38}\ce{Ar} = 36\permil$ at the beginning of the model's time domain - the same as the source magma (see Section \ref{subsec:outgassing}). For $\delta ^{40}\ce{Ar}$, we use the value recorded in meteorite ALH 84001 corresponding to an age of 4.16 Ga \citep{cassata_evidence_2010}, $^{40}\ce{Ar}/^{36}\ce{Ar} = 626$ ($\delta ^{40}\ce{Ar} = -671\permil$). 

We use ALH 84001, which has a crystallization age of 4.16 Ga, to derive the conditions at the start of our model time domain (3.8 Ga). It is possible that the atmosphere evolved during the 300 Myr in between. For example, other models for the evolution of Mars's atmosphere \citep{kurokawa_lower_2018} and additional analysis of ALH 84001 \citep{willett_window_2022} indicate that $\delta^{15}\ce{N}$ and $\delta^{38}\ce{Ar}$ may have been elevated higher than the source magma at the start of our modeled period. To account for this, we also consider an endmember scenario in which enrichment in the heavy isotope occurred prior to 3.8 Ga, where we assume $\delta^{15}\ce{N} = 300\permil$ and $\delta^{38}\ce{Ar} = 1255\permil$.

\subsection{Backward MCMC Search} \label{subsec:backward_mcmc_search}

\begin{figure*} [t!]
\epsscale{1.15}
\plotone{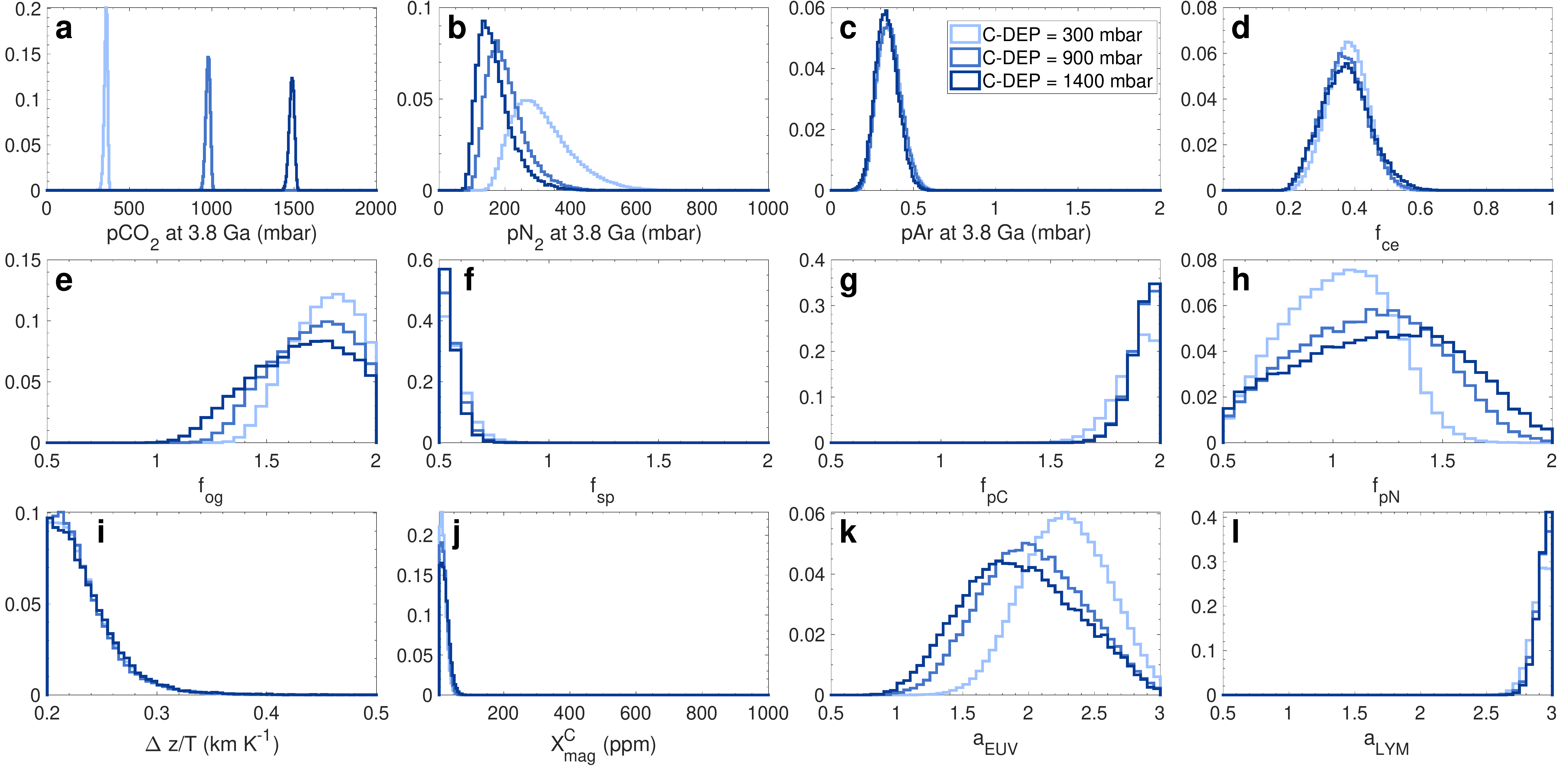}
\caption{Posterior distributions for atmospheric composition at 3.8 Ga and model parameters in the baseline MCMC solutions. The posterior distributions are reported for discrete values of the amount of \ce{CO2} deposited in carbonate minerals before 3.5 Ga, C-DEP = 300, 900, and 1400 mbar. (a) The distributions of \ce{pCO2} at 3.8 Ga have medians (and 95\% confidence intervals) of 357 (334-373), 977 (950-1000), 1486 (1453-1513) mbar, respectively. (b) For \ce{pN2} at 3.8 Ga: 299 (182-522), 192 (119-348), 159 (97-302) mbar. (c) For \ce{pAr} at 3.8 Ga: 0.35 (0.23-0.51), 0.35 (0.23-0.50), 0.34 (0.21-0.48) mbar. (d-l) See Table \ref{table:mcmc_params} for parameter descriptions.}
\label{fig:baseline_posteriors}
\end{figure*}

To explore the parameter space we employ a backward-integrated Markov-Chain Monte Carlo (MCMC) method. On each iteration of the MCMC, we start from the present-day atmospheric partial pressures on Mars and integrate the atmospheric pressure backward in time until 3.8 Ga. This ensures that the predicted atmospheric abundances at 3.8 Ga are consistent with the present-day atmospheric abundances for the given set of parameters. At 3.8 Ga, the isotopic composition of each species is then set according to the source magma or Martian meteorite measurements (see Section \ref{subsec:init_conds}). The partial pressures and isotopic compositions are then integrated forward in time from 3.8 Ga to present, and the modeled present-day isotopic composition is compared to the measured values. The size of the timestep is selected such that the total atmospheric pressure changes by no more than 0.05$\%$. The likelihood function (L) is defined as:
\begin{equation}
    \log L = 
    \sum
    \left(\frac{\delta_{\rm 0,observed} - \delta_{\rm 0,model}}{\sigma_{\rm \delta}}\right)_i^2, \label{eq:likelihood}
\end{equation}

where the sum is over the isotopic compositions tracked in the model ($\delta ^{13}\ce{C}$, $\delta ^{15}\ce{N}$, $\delta ^{38}\ce{Ar}$, $\delta ^{40}\ce{Ar}$) and $\sigma$ is the uncertainty for the present-day isotopic composition of the free reservoir. For each MCMC simulation, two chains are produced starting from parameters chosen independently and randomly within the allowed ranges. These two chains are then tested for convergence using the Gelman-Rubin method \citep{gelman_inference_1992} and if converged, they are combined. Model runs from these chains that reproduce the modern abundances and isotopic composition of all 3 species to within 3-$\sigma$ uncertainty are deemed solutions and used to derive the posterior distributions shown below. The free parameters and their ranges explored in the MCMC search are summarized in Table \ref{table:mcmc_params}. The extended range MCMC is the same as the baseline MCMC but parameters are allowed to vary in wider ranges to prevent biasing. We also consider several variant scenarios, described above, and summarized in Table \ref{table:variants}. Variant scenarios are considered one by one, while all other aspects of the model follow the baseline.

\section{Results} \label{sec:results}

We systematically explore the free parameters in our model using the Markov Chain Monte Carlo (MCMC) method. This method allows us to find populations of model solutions that are consistent with the modern Martian atmosphere. With these solutions, we generate posterior probability distributions that constrain model parameters and the composition of the ancient atmosphere. A solution is defined as a model run that reproduces the modern isotopic composition of all 3 species (\ce{CO2}, \ce{N2}, and \ce{Ar}) in the free reservoir to within 3$\sigma$ of their modern values. The modern abundances of the species are reproduced in all model evolutions because of the backtracking method we employ (See Section \ref{subsec:backward_mcmc_search}). We first discuss results from the baseline MCMC, including parameter probability distributions and individual evolutionary scenarios, then we show results from various sensitivity studies.

\subsection{The Baseline Model} \label{subsec:baseline}

\begin{figure*} [t!]
\epsscale{1.15}
\plotone{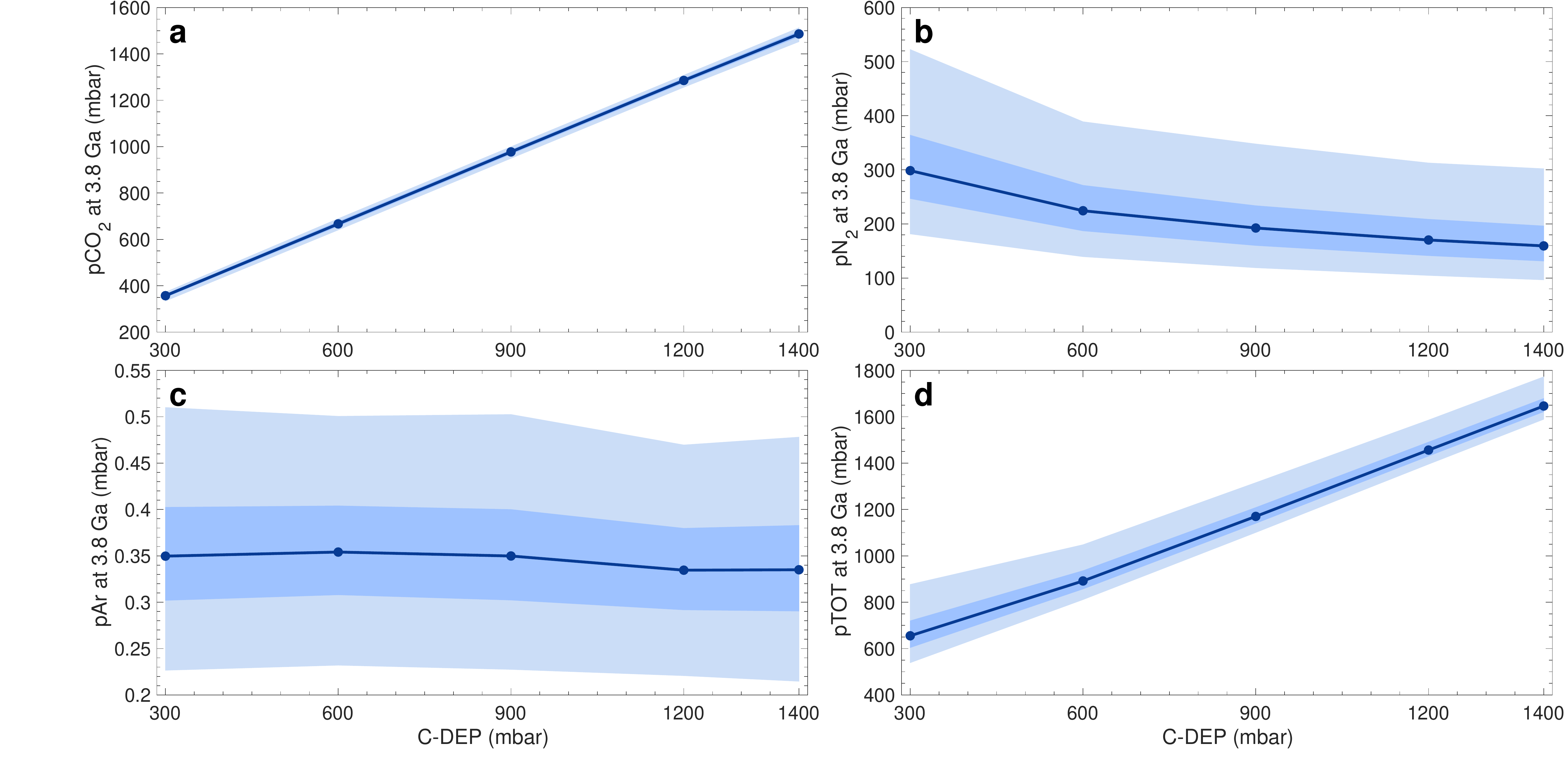}
\caption{Atmospheric composition at 3.8 Ga as a function of early carbonate deposition (C-DEP) in the baseline MCMC. The circular markers are median values from the baseline MCMC posterior distributions. The dark shaded regions are 50\% confidence intervals and the light shaded regions are 95\% confidence intervals. pTOT is defined as the sum of \ce{pCO2}, \ce{pN2}, and \ce{pAr} in a model run.}
\label{fig:baseline_comp}
\end{figure*}

The parameter ranges explored in our baseline MCMC are shown in Table \ref{table:mcmc_params} and the assumptions made are shown in Table \ref{table:variants}. Five different values for the amount of carbonate deposition before 3.5 Ga (C-DEP) were tested: 300, 600, 900, 1200, and 1400 mbar. Each MCMC successfully converged and yielded at least 25,000 solutions. 

The baseline MCMC solutions constrain the composition and size of the Martian atmosphere at 3.8 Ga. Figures \ref{fig:baseline_posteriors}a, b, and c show the posterior distributions for \ce{pCO2}(3.8Ga), \ce{pN2}(3.8Ga), and \ce{pAr}(3.8Ga) as a function of C-DEP in the baseline MCMC solutions. The posterior distributions for \ce{pCO2}(3.8Ga) are sharply peaked and show a strong dependence on C-DEP. We place overall constraints on the abundance of a species by taking the lowest and highest bounds from the 95$\%$ confidence intervals of the three distributions. For \ce{pCO2}(3.8Ga), it is the lower bound from the C-DEP = 300 mbar distribution and upper bound from the C-DEP = 1400 mbar distribution. Thus, \ce{pCO2}(3.8Ga) is constrained to the range 334-1513 mbar. The posterior distributions for \ce{pN2}(3.8Ga) show a weak anti-correlation with C-DEP, and we find that \ce{pN2}(3.8Ga) is constrained to the range 97-522 mbar. The posterior distributions for \ce{pAr}(3.8Ga) are invariant to C-DEP, and we find that \ce{pAr}(3.8Ga) is constrained to the range 0.21-0.51 mbar.

\begin{figure*} [t!]
\epsscale{1.15}
\plotone{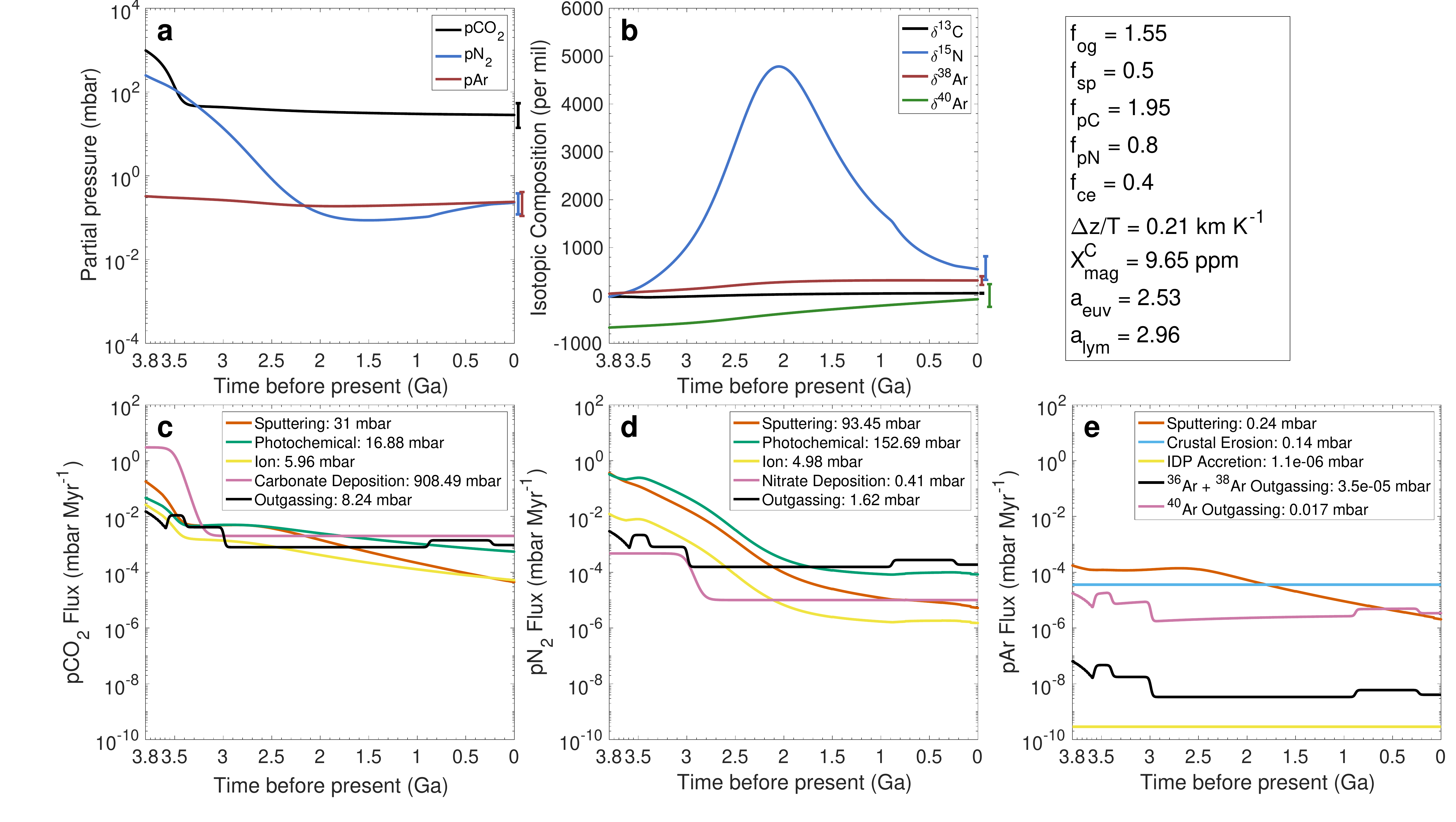}
\caption{Representative model solution from the baseline MCMC. Parameter values for this solution are shown at the top right. \textbf{(a)} Evolution of partial pressures. Modern value ranges are shown by the brackets. At 3.8 Ga, $\ce{pCO2} = 982$ mbar, $\ce{pN2} = 250$ mbar, and $\ce{pAr} = 0.33$ mbar. \textbf{(b)} Evolution of the isotopic composition. Modern values with $3\sigma$ uncertainty are shown by the brackets. \textbf{(c, d, e)} Evolution of the fluxes from sources and sinks of each species. All fluxes are shown as positive values for clarity. In reality, sputtering, photochemical loss, ion loss, carbonate deposition, and nitrate deposition are negative fluxes. The total mass lost or gained over the entire evolution for a given process is shown next to its name in the legends.}
\label{fig:baseline_repcase}
\end{figure*}

C-DEP strongly determines the atmospheric size and composition at 3.8 Ga. Figure \ref{fig:baseline_comp}a shows the linear relationship between C-DEP and \ce{pCO2}(3.8Ga), with minimal spread in the 95$\%$ confidence intervals. As shown in a representative model solution (Figure \ref{fig:baseline_repcase}), carbonate deposition is responsible for over 90$\%$ percent of \ce{CO2} loss, making it more effective than any other source or sink by a factor of $>20$. C-DEP is anti-correlated with \ce{pN2}(3.8Ga) (Figure \ref{fig:baseline_comp}b). This is mainly due to atmospheric escape at early times in the evolution, and is discussed below. Finally, C-DEP has little correlation with \ce{pAr}(3.8Ga) (Figure \ref{fig:baseline_comp}c), and it is a minor component of the ancient atmosphere in all scenarios. Ultimately,  C-DEP has the strongest effect on the size and composition of Mars's atmosphere at 3.8 Ga; but regardless of the value for C-DEP, \ce{pN2}(3.8Ga) is consistently higher than $\sim$100 mbar, with the specific amount depending on C-DEP and the corresponding \ce{CO2} evolution.

\begin{figure*}[t]
\epsscale{1.15}
\plotone{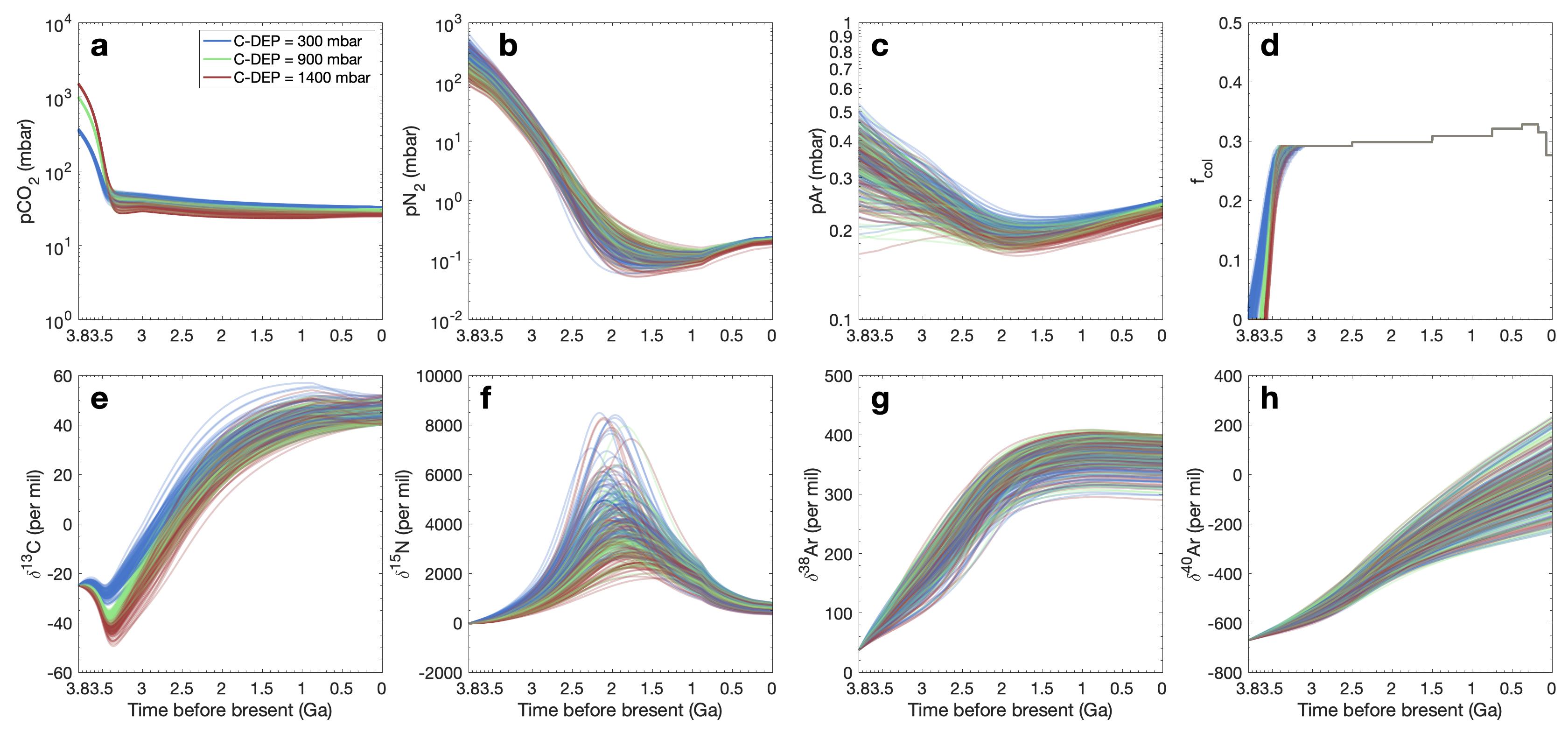}
\caption{100 randomly sampled solutions for each value of C-DEP from the baseline MCMC. The plotted data is slightly transparent for visibility. The solutions are shown with discrete values of the amount of \ce{CO2} deposited in carbonate minerals before 3.5 Ga, C-DEP = 300, 900, and 1400 mbar. Note that panels a, b, and c have a logarithmic X-axis. In panel d, f$_{\rm col}$ is the fraction of time the atmosphere is in the collapsed state for a given timestep (see Section \ref{subsec:freeres_collapse}).}
\label{fig:baseline_sample}
\end{figure*}

A very low concentration of \ce{CO2} in the source magma for volcanic outgassing (X$^C_{\rm mag}$) is preferred in model solutions (Figure \ref{fig:baseline_posteriors}j). X$^C_{\rm mag}$ is allowed to vary in the range 5-1000 ppm, yet it's posterior distribution is pressed against the lower boundary for all values of C-DEP. This behavior can be explained by examining the $\delta^{13}$C evolution. Atmospheric escape does not strongly fractionate carbon in \ce{CO2} over the course of an evolution: photochemical escape is an effective fractionator, but does not occur at a high enough rate, and although sputtering occurs at a higher rate, it is not an effective fractionator. These processes must fractionate \ce{CO2} enough to raise $\delta^{13}$C from -25$\permil$ at 3.8 Ga to 46$\permil$ at present-day despite the continuous addition of volcanically outgassed \ce{CO2} that has $\delta^{13}$C = -25$\permil$. Thus, in order to recreate the modern $\delta^{13}$C, high fractionation from atmospheric escape and low \ce{CO2} volcanic outgassing are required, as shown by the parameter distributions relevant to these processes in Figures \ref{fig:baseline_posteriors}e, g, j, and l. One might ask why must X$^C_{\rm mag}$ be low, and not the total rate of volcanic outgassing instead? The total rate of volcanic outgassing cannot be lowered via the outgassing multiplier ($f_{\rm og}$) because a non-negligible amount of recently outgassed \ce{N2} is required to recreate the modern $\delta^{15}$N value. Lowering the total outgassing rate would prevent this constraint from being satisfied, which forces X$^C_{\rm mag}$ to be low instead. 

\ce{N2} follows a dynamical track evolution in all solutions found in the baseline MCMC. The dynamical track evolution, first identified in a model containing only \ce{N2} \citep{hu_nitrogen-rich_2022}, is an evolutionary track for \ce{pN2} characterized by high abundance at 3.8 Ga and a gradual descent to the modern partial pressure. Figure \ref{fig:baseline_sample} shows a random sample of \ce{N2} evolutionary tracks from the baseline MCMC solutions in which \ce{N2} descends to the modern pressure around 2 Ga. This late descent is characteristic of the dynamical track solutions, and the population shown here is directly comparable to the dynamical track solutions in \citet{hu_nitrogen-rich_2022}, their Figure 3a. The dynamical track solution is fundamentally due to the decoupling of the \ce{N2} sputtering rate from the \ce{N2} mixing ratio when the atmospheric abundances of \ce{N2} and \ce{CO2} are comparable \citep{hu_nitrogen-rich_2022}. This behavior is observed in Figure \ref{fig:baseline_repcase}d when the \ce{N2} sputtering rate is unchanged by the rapid \ce{CO2} decline before 3.5 Ga, as opposed to photochemical and ion escape. Moreover, the dynamical track solutions are favored when parameters are kept closest to their nominal values because escape rates are not high enough to quickly drive \ce{pN2} down to its modern value. Dynamical track solutions are key for constraining the ancient atmospheric composition because they require a unique value of \ce{pN2}(3.8Ga).

\ce{pN2}(3.8Ga) is weakly anti-correlated with \ce{pCO2}(3.8Ga) because a high partial pressure of \ce{CO2} reduces the atmospheric escape rate of \ce{N2}. The mixing ratio of \ce{N2} in the atmosphere is lower when there is a larger amount of \ce{CO2}. The atmospheric escape rate of \ce{N2} depends directly on its mixing ratio. So, a larger value of \ce{pCO2} causes a lower \ce{N2} mixing ratio and thus a lower atmospheric escape rate of \ce{N2}. Because atmospheric escape is the dominant sink for \ce{N2}, a lower escape rate means there must be less \ce{pN2} at 3.8 Ga. This effect is especially important at early times when atmospheric \ce{CO2} has not been significantly lost to carbonate deposition, atmospheric collapse hasn't occurred yet, and the early sun was emitting more EUV radiation which drives \ce{N2} escape.


The evolution of \ce{N2} is characterized by the relationship between early atmospheric escape and recent volcanic outgassing. Early atmospheric escape from sputtering and photochemical loss strongly fractionates \ce{N2} and raises $\delta^{15}$N as \ce{pN2} is driven to its low, near-modern value. With low \ce{pN2}, atmospheric escape processes become less efficient and the \ce{N2} introduced from Mars's interior at recent times has a stronger influence on the $\delta^{15}$N, which lowers it to the modern value. This behavior is also what drives the posterior distribution of $\Delta z/T$ to its lower boundary (Figure \ref{fig:baseline_posteriors}i). A higher $\Delta z/T$ would enhance the fractionation from all atmospheric escape processes. A higher recent outgassing rate would then be required in order to offset the enhanced fractionation. However, a larger volcanic outgassing rate cannot be employed because it would introduce too much \ce{N2} into the atmosphere and raise \ce{pN2} above the modern value. Thus, a low $\Delta z/T$ value is preferred in the posterior distributions.

The evolution of argon constrains the overall sputtering rate and strengthens the constraints placed by other species. Sputtering is the only sink for all argon isotopes, and the only process that can significantly increase $\delta^{38}$Ar. Thus, differences in the sputtering rate have the strongest effect on the argon evolution. The posterior distribution preference for a low sputtering rate (Figure \ref{fig:baseline_posteriors}f) is a consequence of this. A higher sputtering rate would require a larger value of \ce{pAr}(3.8Ga), and the sputtering required to reach the modern \ce{pAr} value would cause over-fractionation of $\delta^{38}$Ar. This cannot be compensated by increased Ar outgassing with a larger $f_{\rm og}$ because it would force X$^C_{\rm mag}$ to be lower than the allowed range (see above). Because sputtering is a process that effects all three species, argon is thus constraining the entire evolution.

\subsection{Sensitivity Tests}

In this section we discuss the results from testing assumptions made in the baseline model. This includes extending the explored parameter ranges, changing the volcanic outgassing chronology, changing the carbonate deposition scenario, removing the treatment of atmospheric collapse, and exploring a scenario where the delta values at 3.8 Ga are higher due to potential fractionation before the modeled period.

\subsubsection{Extended Parameter Ranges}

\begin{figure*}[t!]
\epsscale{1.15}
\plotone{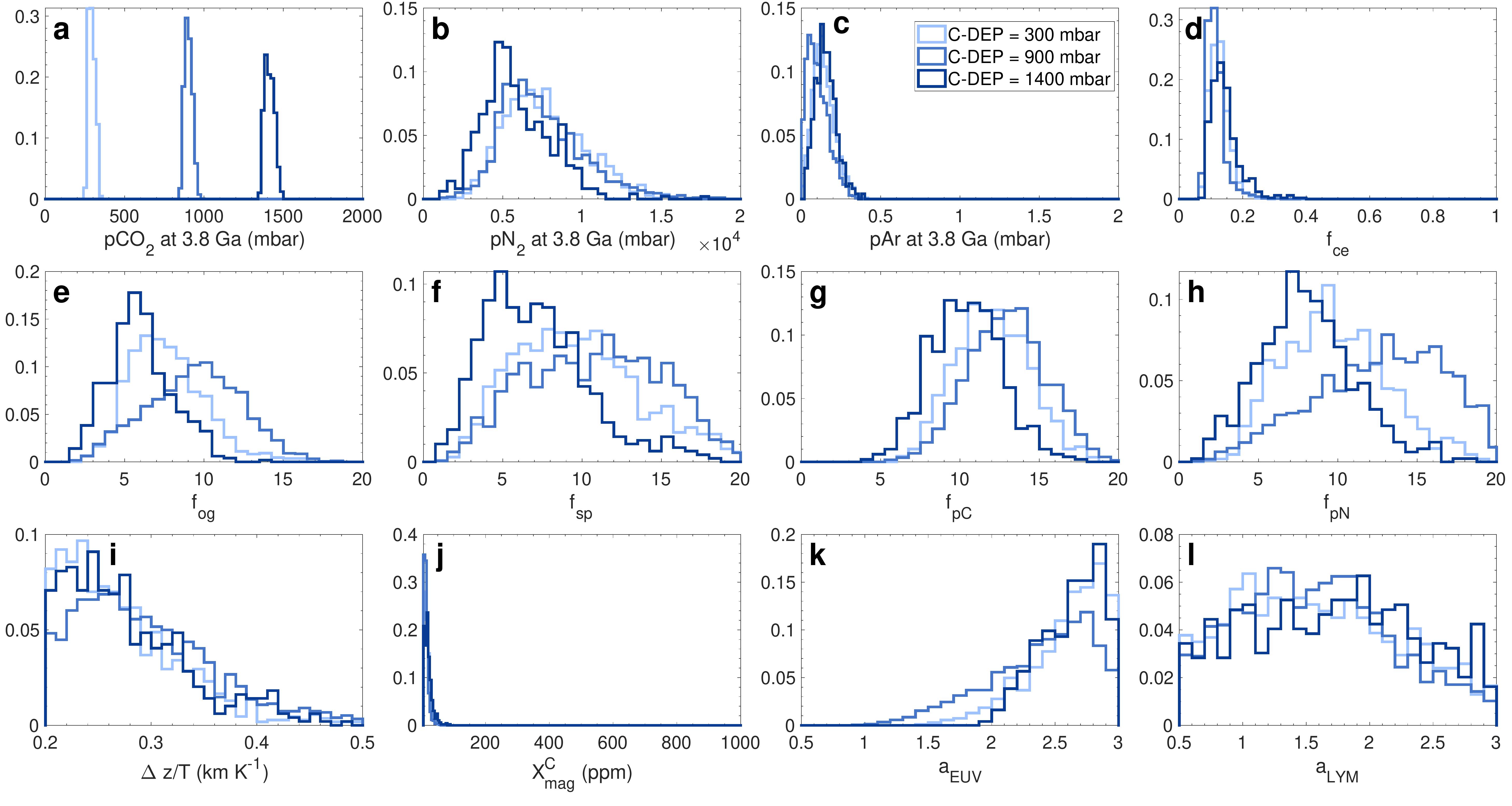}
\caption{Posterior distributions for atmospheric composition at 3.8 Ga and model parameters in the extended range MCMC solutions. The posterior distributions are reported for discrete values of the amount of \ce{CO2} deposited in carbonate minerals before 3.5 Ga, C-DEP = 300, 900, and 1400 mbar. (a) The distributions of \ce{pCO2} at 3.8 Ga have medians (and 95\% confidence intervals) of 290 (260-334), 900 (861-949), 1412 (1364-1475) mbar, respectively. (b) For \ce{pN2} at 3.8 Ga: 7.33 (3.71-13.02), 7.00 (3.43-13.75), 5.45 (2.35-10.12) bar. Note the large scale of the X-axis. (c) For \ce{pAr} at 3.8 Ga: 0.134 (0.028-0.300), 0.094 (0.009-0.264), 0.149 (0.040-0.311) mbar. (d-i) See Table \ref{table:mcmc_params} for parameter descriptions.}
\label{fig:extended_posteriors}
\end{figure*}

\begin{figure}
\epsscale{1.15}
\plotone{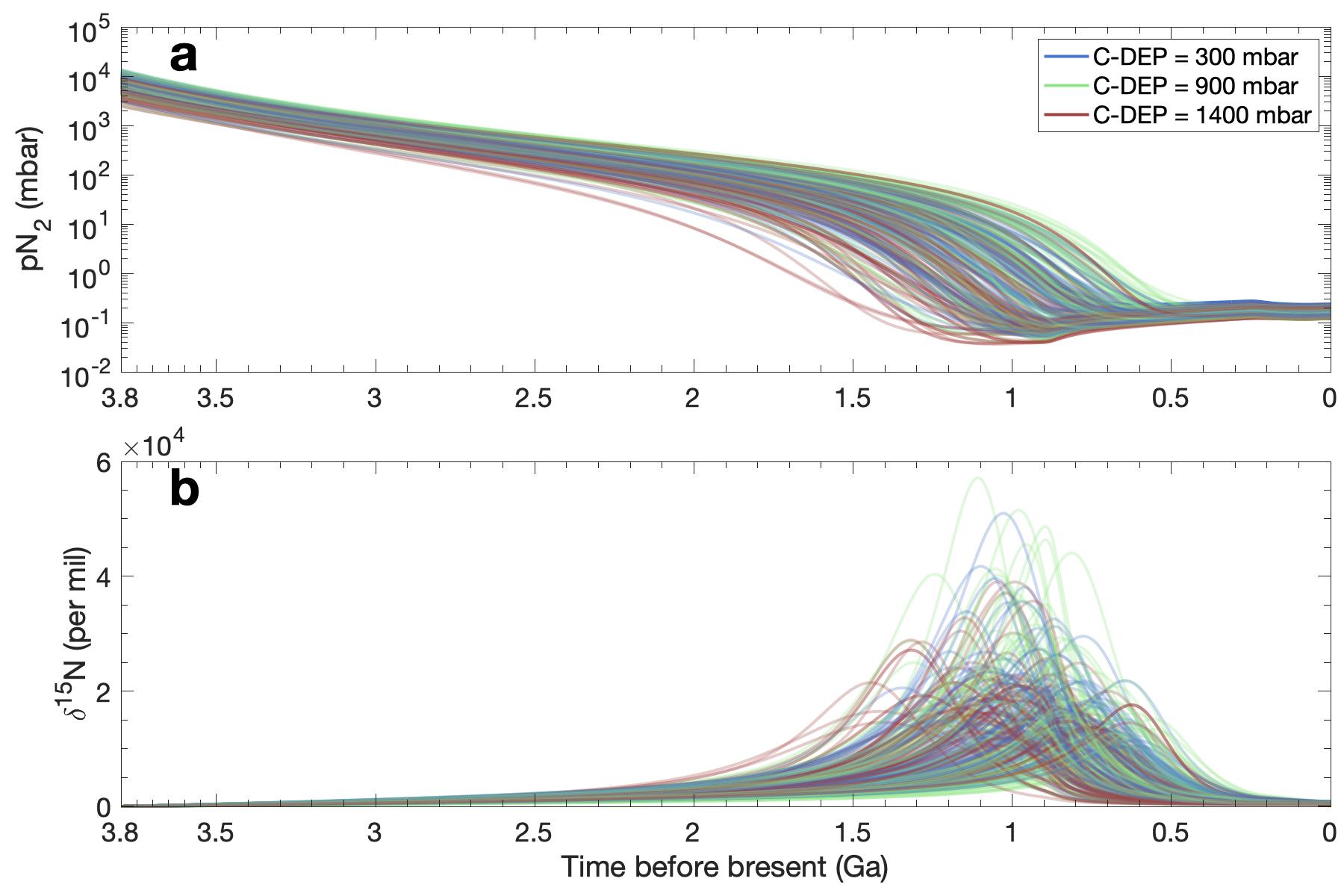}
\caption{Evolution of \ce{N2} in 100 randomly sampled solutions from the extended range MCMC. The plotted data is slightly transparent for visibility. The solutions are shown with discrete values of the amount of \ce{CO2} deposited in carbonate minerals before 3.5 Ga, C-DEP = 300, 900, and 1400 mbar.}
\label{fig:extended_n2sample}
\end{figure}

\begin{figure*} [t!]
\epsscale{1.15}
\plotone{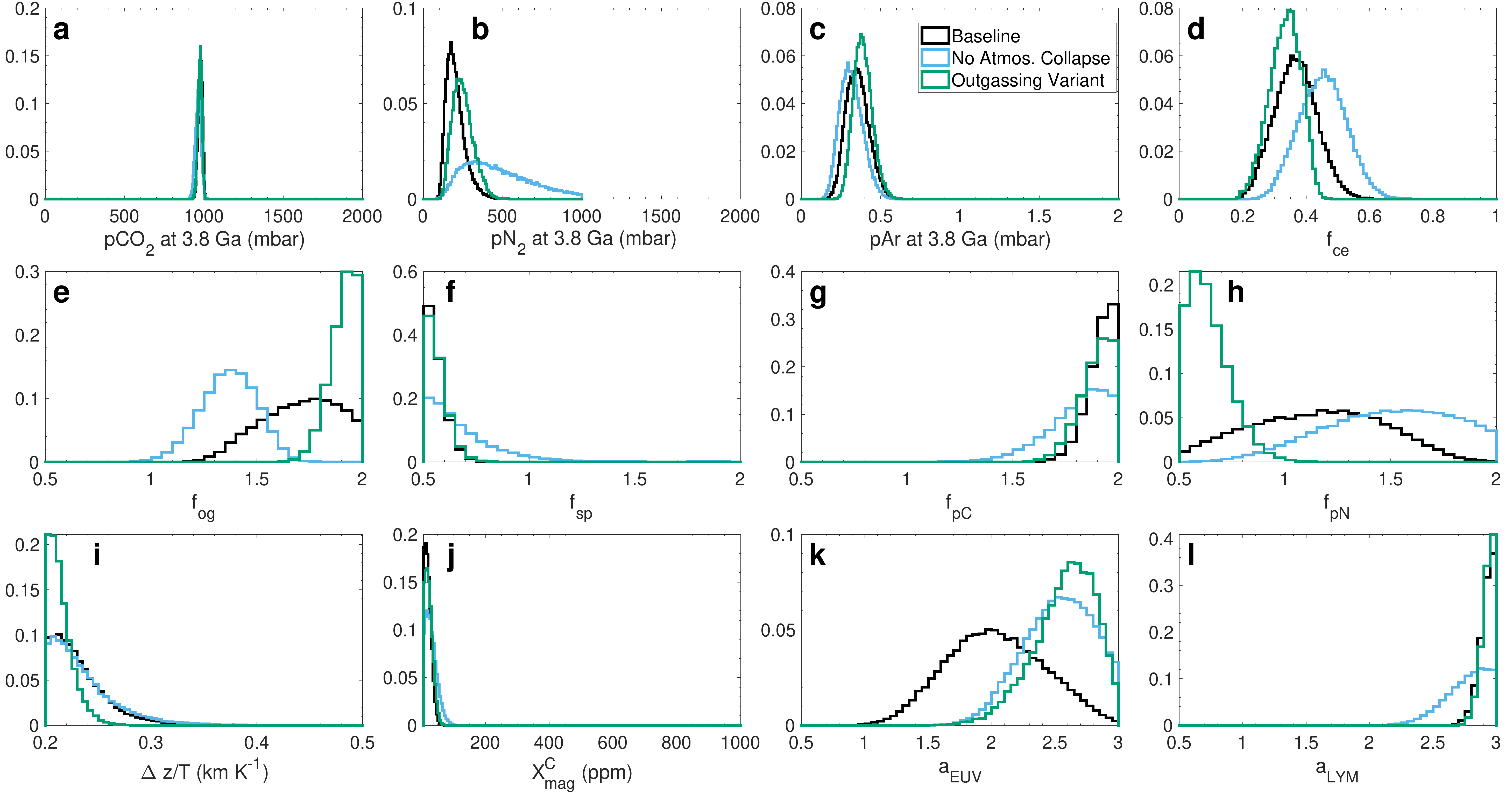}
\caption{Posterior distributions for atmospheric composition at 3.8 Ga and model parameters in the variant MCMC solutions. All posteriors have an amount of carbonate deposition before 3.5 Ga (C-DEP) of 900 mbar. More information on the variant scenarios is shown in Table \ref{table:variants}.}
\label{fig:variant_posteriors}
\end{figure*}

Here we discuss results from the extended range MCMC (Table \ref{table:mcmc_params}), where model free parameters are allowed to vary in a larger range of values than the baseline MCMC. We explore this scenario because, although the boundaries we impose in the baseline MCMC are physically reasonable, they may influence the results of the MCMC. The ranges for $f_{\rm og}, f_{\rm sp}, f_{\rm pC},$ and $f_{\rm pN}$ are extended from 0.5-2 to 0.01-20. The other parameter ranges are unchanged.

The solutions in the extended range MCMC are similar to the solutions in the baseline MCMC, but they allow massive \ce{pN2} atmospheres at 3.8 Ga. The posterior distribution in the extended range (Figure \ref{fig:extended_posteriors}b) shows that \ce{pN2}(3.8Ga) peaks around 5 bar but can be as high as 20 bar. The posterior distributions in the extended range also show that all 4 multipliers are taking on values much higher than those in the baseline MCMC (Figure \ref{fig:extended_posteriors}e, f, g, and h). The increased atmospheric escape requires a large \ce{N2} atmosphere at 3.8 Ga in order to recreate the modern \ce{pN2} value it requires a large rate of volcanic outgassing rate to ensure that recreate the modern $\delta^{15}$N value. As shown in Figure \ref{fig:extended_n2sample}, these solutions follow a very similar trajectory to the baseline solutions, and \ce{pN2} descends to its modern value late in the evolution. Other important aspects of the extended range solutions are the same as the baseline solutions: \ce{pCO2}(3.8Ga) is still tightly constrained by C-DEP, \ce{pAr}(3.8Ga) is still invariant to C-DEP albeit slightly lower, and the posterior distribution for X$^C_{\rm mag}$ is still at the lower end of its boundary.

\subsubsection{Volcanism Profile}

\begin{figure*} [t!]
\epsscale{1.15}
\plotone{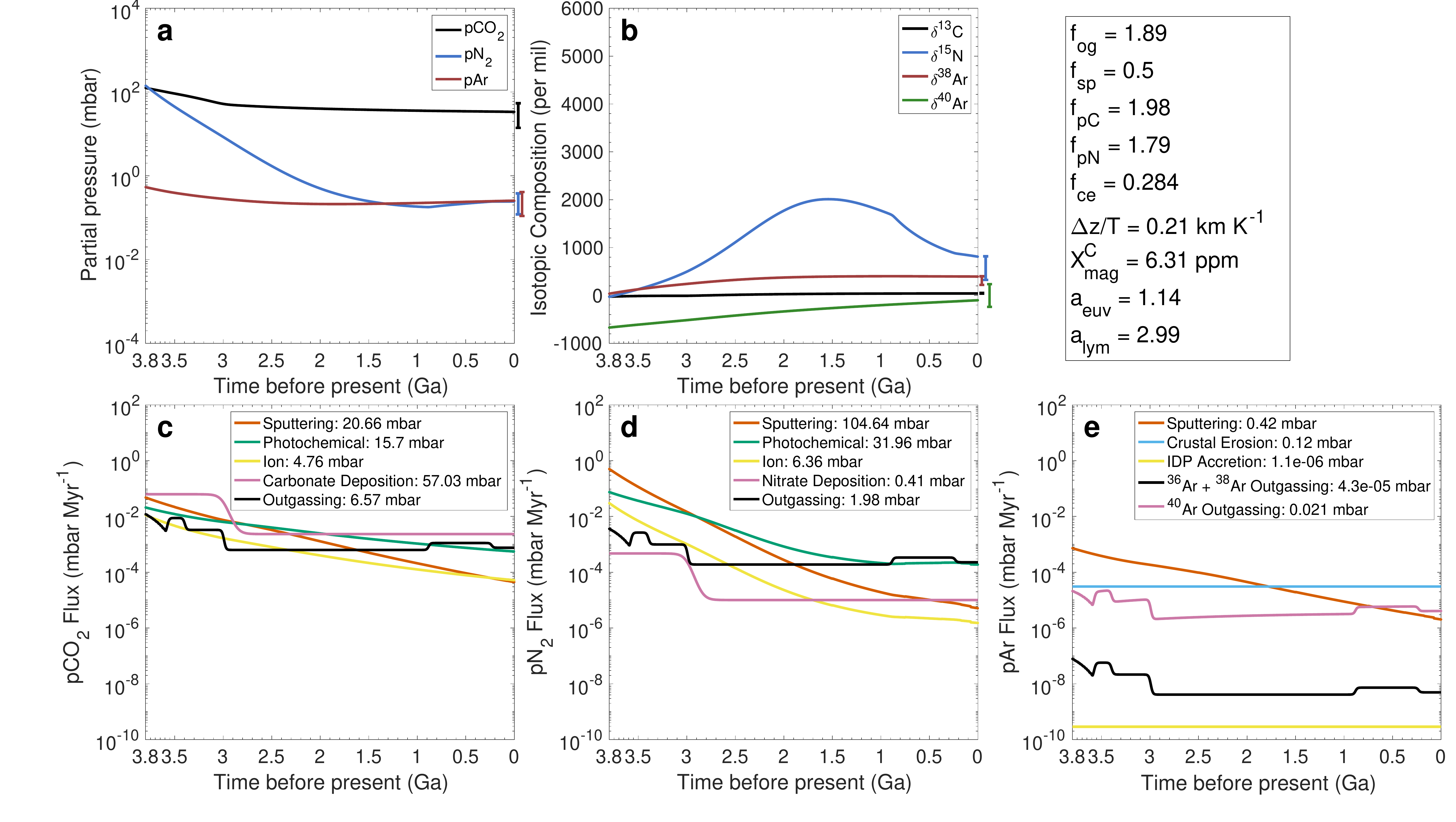}
\caption{Representative model solution from the variant case where carbonate deposition occurred in shallow subsurface aquifers (SSA). This case has 50 mbar of carbonate deposition before 3 Ga. Parameter values for this solution are shown at the top right. \textbf{(a)} Evolution of partial pressures. Modern value ranges are shown by the brackets. At 3.8 Ga, $\ce{pCO2} = 125$ mbar, $\ce{pN2} = 142$ mbar, and $\ce{pAr} = 0.54$ mbar. \textbf{(b)} Evolution of the isotopic composition. Modern values with $3\sigma$ uncertainty are shown by the brackets. \textbf{(c, d, e)} Evolution of the fluxes from sources and sinks of each species. All fluxes are shown as positive values for clarity. In reality, sputtering, photochemical loss, ion loss, carbonate deposition, and nitrate deposition are negative fluxes. The total mass lost or gained over the entire evolution for a given process is shown next to its name in the legends.}
\label{fig:SSAcarb_repcase}
\end{figure*}

To test the model sensitivity to the rate of volcanic outassing over time, we run the baseline MCMC with the variant volcanism profile derived from the crater chronology of \citet{kallenbach_marsmoon_2001} instead of \citet{hartmann_martian_2005}. The two chronologies are similar at the start of the model time domain, but the primary difference is that the variant chronology has less volcanic outgassing in the most recent billion years. This difference primarily affects \ce{N2} because the baseline solutions rely on recent volcanic outgassing to lower $\delta^{15}$N from its escape-induced peak to the modern value. To compensate for this, the MCMC posterior distributions show enhanced total volcanic outgassing and reduced photochemical escape of nitrogen (Figure \ref{fig:variant_posteriors}). This parameter change effectively mitigates the difference in volcanic rate profiles, as the posteriors for atmospheric composition at 3.8 Ga and other parameters are very similar. Finally, it is interesting to note that the posterior distribution for the concentration of \ce{CO2} in source magma (X$^C_{\rm mag}$) is unchanged, indicating that this result is robust to changes in the net outgassing rate over time.

\subsubsection{Atmospheric Collapse}

Atmospheric collapse causes enhanced escape rates of \ce{N2} and \ce{Ar}. When the atmosphere collapses in the baseline scenario, the atmospheric \ce{pCO2} used to calculate mixing ratios drops from potentially hundreds of mbars to 6 mbar. This drastically enhances the mixing ratios of \ce{N2} and \ce{Ar}, which in turn enhances their atmospheric escape rates. In the baseline MCMC solutions with C-DEP = 900 mbar, the average model run had a collapsed atmosphere 28$\%$ of the time. A sample of the collapse probability over time is shown in Figure \ref{fig:baseline_sample}d. In all cases, the probability becomes non-zero after a few hundred Myrs as \ce{pCO2} declines due to carbonate deposition. After \ce{pCO2} becomes sufficiently low, all samples tend to follow the same path because the obliquity of Mars -- not the size of the atmosphere -- is the limiting factor for collapse.

Posterior distributions in the MCMC are adjusted to compensate for the effects of ignoring atmospheric collapse (Figure \ref{fig:variant_posteriors}). The MCMC posterior distributions show that parameters related to photochemical escape of \ce{N2} are preferred to be higher than in the baseline scenario. With these parameter values, the \ce{N2} escape rate is raised closer to what it would be if atmospheric collapse was included. Solutions with higher \ce{pN2} at 3.8 Ga then become available because the \ce{N2} photochemical escape rate is also higher at early times, when \ce{pCO2} is high enough so that the atmosphere would not be collapsed in the baseline scenario anyway. So, although ignoring atmospheric collapse means a lower \ce{N2} escape rate, compensating for this effect creates solutions with even higher \ce{pN2}(3.8Ga). Additionally, the posterior distribution for the outgassing multiplier peaks at a lower value than the baseline and it is independent of the parameter boundaries. The evolution of \ce{Ar} is responsible for this. Atmospheric escape of \ce{Ar} is reduced for the same reasons as \ce{N2}; however, \ce{Ar} is only removed via sputtering, which must remain low to avoid over-fractionation (see Section \ref{subsec:baseline}). Ignoring atmospheric collapse then results in a less effective total sink for \ce{Ar} that cannot be compensated by increasing sputtering. Thus, volcanic outgassing is instead reduced to ensure that the modern pressure is recreated.

\subsubsection{Carbonate Deposition Scenario}

No model solutions are found with 100 or more mbar of carbonate deposition in the shallow subsurface aquifer (SSA) scenario. We tested model scenarios with 50, 100, 150, 200, and 250 mbar of SSA carbonate deposition in addition to the standard model runs with 300, 900, and 1400 mbar carbonate deposition. Model solutions are only found in the scenario with 50 mbar of SSA carbonate deposition - a representative case is shown in Figure \ref{fig:SSAcarb_repcase}. The atmosphere at 3.8 Ga in this representative case has 125 mbar \ce{CO2} and \ce{142} mbar \ce{N2}. The low amount of carbonate deposition causes less \ce{CO2} at the start of the model because it is the main \ce{CO2} sink.

Enhanced \ce{CO2} fractionation in the SSA scenario causes low carbonate deposition and low \ce{pCO2}(3.8Ga). In a typical model run in this scenario, solutions are difficult to find because the modeled $\delta^{13}$C almost always ends up lower than the modern value. This is due to the fact that deposition in the SSA scenario fractionates carbon more strongly than deposition in open water systems (OWS), which is the baseline. The fractionation factor for deposition in SSA is 1.06, whereas in OWS it is 1.01. With more than 100 mbar of deposition, the atmosphere becomes significantly more enriched in the light isotope in the SSA scenario as opposed to the OWS scenario. In the baseline OWS model, atmospheric escape of carbon is able to offset this process and fractionate the atmosphere in the other direction enough to bring it to the modern value. Our MCMC search finds no scenarios where atmospheric escape can offset the enhanced fractionation from $\geq$100 mbar of carbonate deposition in SSA.  Thus, there must have been less than 100 mbar of carbonate deposition if it was entirely in SSA. Because carbonate deposition is the dominant atmospheric \ce{CO2} sink, this endmember scenario would require a small ancient atmosphere. In reality, carbonate deposition may have occurred in a combination of both scenarios. In general, the more SSA deposition occurred, the smaller the ancient atmosphere should be.

\subsubsection{Isotopic Composition at 3.8 Ga}

No model solutions were found in the scenario with elevated $\delta^{15}$N and $\delta^{38}$Ar values at 3.8 Ga. This scenario is presumed to be possible if there was enhanced atmospheric escape and fractionation of Ar and N before our model time domain. The modern $\delta^{15}$N can still be recreated in this model scenario, as recent outgassing is always able to restore $\delta^{15}$N enough to reach the modern value. No model solutions were found because the modern $\delta^{38}$Ar cannot be reproduced in this scenario, as the model produces a value that is too high. Even though $\delta^{38}$Ar in outgassed material is 36$\permil$, the argon concentration is too low in the outgassed material to restore $\delta^{38}$Ar to the modern value. The modern $\delta^{38}$Ar still cannot be recreated when the MCMC explored an increased volcanic outgassing multiplier and a decreased sputtering multiplier, both of which should help lower $\delta^{38}$Ar. Thus, if $\delta^{38}$Ar was enhanced at 3.8 Ga, our models suggest that there must have been some other process affecting the subsequent $\delta^{38}$Ar evolution, such as a cometary impact \citep{kurokawa_lower_2018}.

\section{Discussion} \label{sec:disc}

We have presented the first coupled, self-consistent simulations for the evolution of \ce{CO2}, \ce{N2}, and \ce{Ar} that reproduce the modern abundances and isotopic composition of the Martian atmosphere. Our results provide insights into the ancient atmospheric composition and its subsequent evolution to the modern day state. Many of the critical constraints we've found arise from modeling multiple atmospheric species at once and would not be found in a model that includes just one species. This work is a comprehensive analysis of Mars's atmospheric evolution and we advocate for more multi-species isotope analysis in future studies. Below we discuss the emerging picture of the ancient atmosphere, comparison with other studies, and model validity.

\subsection{A \ce{CO2}-\ce{N2}-\ce{H2} Atmosphere on Ancient Mars}

Our evolutionary models indicate that a \ce{CO2}-\ce{N2}-\ce{H2} ancient atmosphere is likely to exist on ancient Mars. This multi-component atmosphere provides a promising path to explain the geologic evidence for liquid-water activities in the past.

First, our models indicate that a large amount of \ce{N2} is consistent with a large amount of \ce{CO2} in the ancient atmosphere. This is fundamentally due to the dominance of the dynamical track \ce{N2} solutions, first discovered in a similar model that focused on the evolution of \ce{N2} \citep{hu_nitrogen-rich_2022}. We find that these dynamical track solutions are the only \ce{N2} solution in the fully coupled model, and the steady state solutions have disappeared. The dynamical track solutions are characterized by a large \ce{pN2}(3.8Ga) that gradually descends to the modern day value, as opposed to the steady state solutions which feature rapid decline of \ce{N2} early in the evolution. Importantly, each dynamical track solution has a unique value for \ce{pN2}(3.8Ga), so we are able to reconstruct the ancient atmosphere. We find that Mars's atmosphere contained 0.1-0.5 bar \ce{N2} with 0.3-1.5 bar \ce{CO2} at 3.8 Ga for the explored range of carbonate deposition in our model. As suggested in \citet{hu_nitrogen-rich_2022}, an atmosphere with these relative abundances of C and N could have been formed by a late veneer of primitive bodies such as comets that have low C/N ratios \citep{bergin_tracing_2015}. In any case, our models uniformly suggest that \ce{N2} is a non-negligible part of the ancient atmosphere and should thus be considered for climatic effects and atmospheric chemistry.

Second, we find that recreating the modern \ce{pCO2} and $\delta^{13}$C requires a carbon content in the source magma that indicates a reduced Martian mantle. Across all of the solutions found in every MCMC and variant, the concentration of \ce{CO2} in the source magma for volcanic outgassing ($X^C_{\rm mag}$) was preferred to be at the very bottom of the allowed parameter range.  The lower bound on $X^C_{\rm mag}$ of 5 ppm corresponds to a source magma oxygen fugacity one log unit below the iron-w{\"u}stite buffer, IW-1 \citep{hirschmann_ventilation_2008}. The highest value of $X^C_{\rm mag}$ that yields solutions is around 50 ppm, which corresponds to a source magma oxygen fugacity of IW. If we assume that $X^C_{\rm mag}$ directly reflects the magma oxygen fugacity, then our models require a reduced martian mantle of IW or lower. While modern Mars's interior is thought to be at an oxygen fugacity of IW+1 \citep{grott_volcanic_2011}, analysis of the oldest Martian meteorite, ALH84001, indicates that early Mars's interior could have had an oxygen fugacity as low as IW-1 \citep{warren_siderophile_1996}. Note that this insight arises from the joint analysis of the evolution of \ce{CO2} and \ce{N2}, highlighting the importance of including multiple species in the model.

We can estimate the \ce{H2} volcanic outgassing flux implied by the low mantle oxygen fugacity. We consider the magma to degas at typical outgassing temperature and pressure (1450 K, 5 bar) \citep{holland_chemical_1984}. The \ce{H2}:\ce{H2O} outgassing ratio controlled by the reaction \ce{2 H2O <->[K1] 2 H2 + O2} is:
\begin{equation}
    \frac{P_{\ce{H2}}}{P_{\ce{H2O}}} = \left(\frac{K_1}{f_{\ce{O2}}}\right)^{0.5},
    \label{eq:h2}
\end{equation}
where $P_{\ce{H2}}$ is the outgassed \ce{H2}, $P_{\ce{H2O}}$ is the outgassed \ce{H2O}, $K_1$ is the equilibrium constant of the reaction ($1.80 \times 10^{-12}$ atm) \citep{ramirez_warming_2014}, and $f_{\ce{O2}}$ is the oxygen fugacity of the source magma. At the typical outgassing temperature, an oxygen fugacity at IW corresponds to $f_{\ce{O2}} = 10^{-12.2}$ atm and \ce{H2}:\ce{H2O} = 1.69. For oxygen fugacity at IW-1, $f_{\ce{O2}} = 10^{-13.2}$ atm and \ce{H2}:\ce{H2O} = 5.33. There is no direct meteoric evidence constraining the Mars mantle water content during our highest modeled outgassing rates, 3.6-3.8 Ga. \citet{brasser_formation_2013} estimates the upper bound of the early Mars mantle water content is 2000 ppm by mass based on planet formation models. Other Mars interior modeling studies typically assume 100 ppm by mass as a baseline case \citep{morschhauser_crustal_2011,liu_impact-melt_2018}, and explore up to 1000 ppm by mass \citep{scheller_long-term_2021}. Thus, we explore an \ce{H2O} concentration in the source magma of $100-2000$ ppm by mass. This encompasses the range of water content in Earth magma, which is typically 500-1000 ppm by mass \citep{morbidelli_source_2000}. For simplicity, we assume complete degassing of the \ce{H2O} in magma, as either \ce{H2O} or \ce{H2}, with the molar ratios calculated above. Because the \ce{H2O} content in the magma will decline as it is degassed, our estimates can be considered an upper limit given our assumptions.

\begin{figure} 
\epsscale{1.25}
\plotone{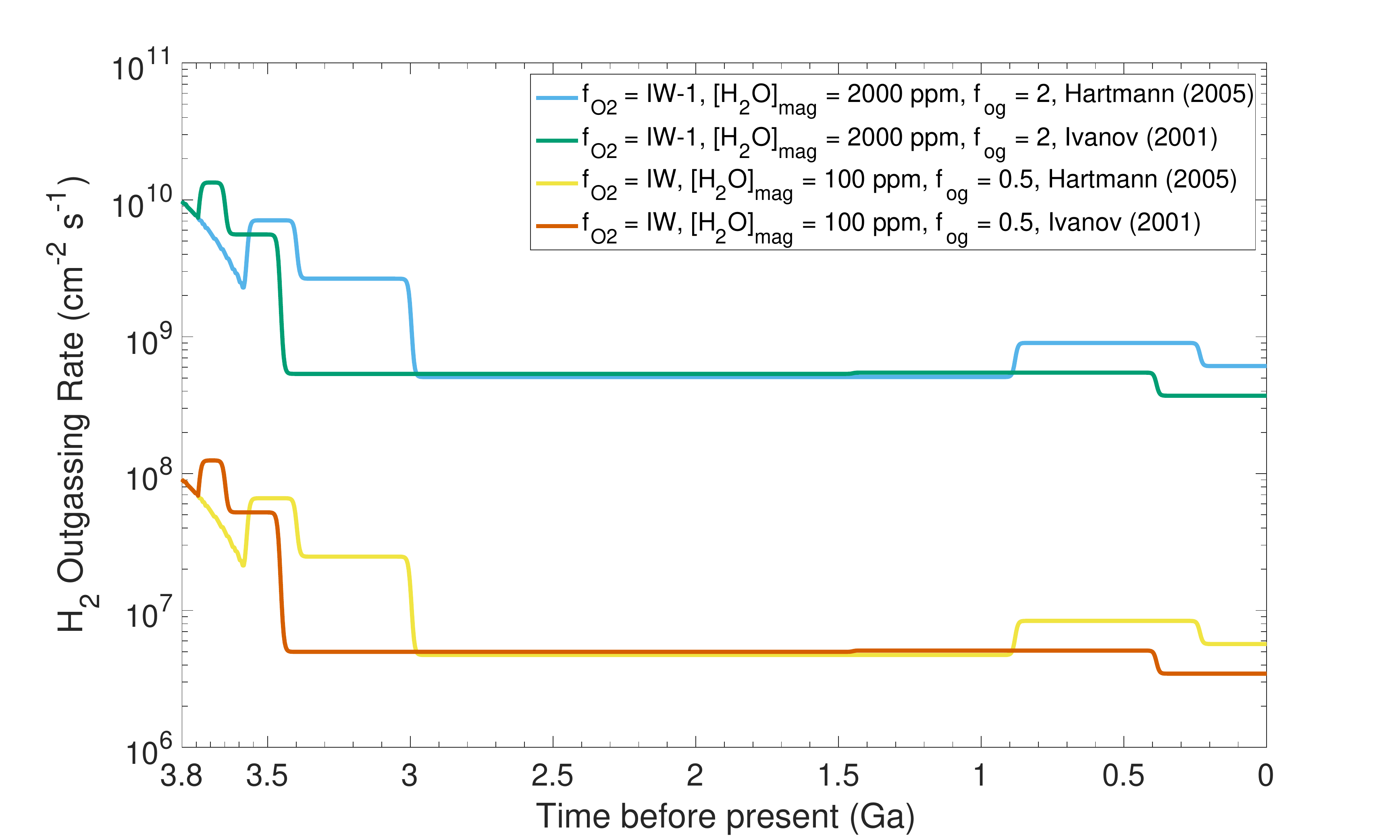}
\caption{Estimated volcanic outgassing rates of \ce{H2}. The conditions that yield maximum outgassing rates are oxygen fugacity at IW-1, \ce{H2O} concentration in the magma of 2000 ppm, and outgassing multiplier of 2. The conditions that yield minimum outgassing rates are oxygen fugacity at IW, \ce{H2O} concentration in the magma of 100 ppm, and outgassing multiplier of 0.5. We test the volcanic profile derived for the baseline case \citep{hartmann_martian_2005} and the variant \citep{kallenbach_marsmoon_2001}.}
\label{fig:h2_rates}
\end{figure}

Our estimate indicates that the rate of \ce{H2} outgassing is likely too low to support a substantial \ce{H2} atmosphere on its own, despite the low oxygen fugacity. We find the \ce{H2} concentration in the magma is 7-94 ppm by mass. We combine the calculated \ce{H2} concentrations with our model crustal production rate to derive the \ce{H2} outgassing rate (Figure \ref{fig:h2_rates}). We consider the range for the volcanic outgassing multiplier employed in our baseline MCMC search (0.5-2) and test both the baseline profile and the variant. The maximum rate of \ce{H2} volcanic outgassing implied by our modeling results is 1.34$\times10^{10}$ cm$^{-2}$ s$^{-1}$, occurring with oxygen fugacity of IW-1, 2000 ppm \ce{H2O} in the magma, volcanic outgassing multiplier of 2, and the variant \citep{kallenbach_marsmoon_2001} volcanic profile. Note that this is well below previous estimates of early Mars's \ce{H2} outgassing rates \citep{ramirez_warming_2014,batalha_testing_2015}, which were derived by scaling the modern Earth's outgassing rate. At least $\sim$3-5\% \ce{H2} is needed to generate substantial warming on ancient Mars for the maximum amount of atmospheric \ce{CO2} we predict \citep{ramirez_warmer_2017,wordsworth_transient_2017,turbet_measurements_2020}. The \ce{H2} outgassing flux we estimate is over an order of magnitude too low to support a 5\% \ce{H2} atmosphere against atmospheric escape \citep{batalha_testing_2015}. Thus, if \ce{H2} was a substantial part of the ancient atmosphere, it must have come from sources other than volcanic outgassing. Crustal hydration and sequestration of water \citep{scheller_long-term_2021} may provide the required \ce{H2} flux (Adams et al., in prep.). 

The atmospheric \ce{N2} abundances we predict can provide surface warming on the order of 10 K on ancient Mars. \citet{von_paris_n2-associated_2013} show that \ce{N2}-\ce{N2} collisional induced absorption (CIA) and pressure broadening of \ce{CO2} absorption lines can cause substantial warming. In our model scenario with C-DEP = 900 mbar, the atmosphere contains $\sim$1 bar \ce{CO2} and 200-400 mbar \ce{N2}. According to \citet{von_paris_n2-associated_2013}, this atmosphere would provide 5-10 K surface warming from \ce{N2}-\ce{N2} CIA and pressure broadening, with respect to an atmosphere with the same amount of \ce{CO2} but without \ce{N2}. Additionally, \ce{N2}-\ce{H2} CIA may contribute minor warming up to 1-2 K for the \ce{N2} levels we predict \citep{wordsworth_hydrogen-nitrogen_2013, ramirez_warming_2014}.

Putting the arguments above together, we suggest that a \ce{CO2}-\ce{N2}-\ce{H2} ancient atmosphere would be consistent with Mars's geochemical evolution, and it provides a promising path forward for explaining the evidence of ancient surface liquid-water. Previous studies have explored a \ce{CO2}-\ce{H2} greenhouse on early Mars to produce the necessary warming to explain the geologic evidence \citep[e.g.,][]{ramirez_warming_2014, ramirez_climate_2020,wordsworth_coupled_2021}. In order to produce even transient warming and melting in these studies, large \ce{CO2} atmospheres and large \ce{H2} sources to the atmosphere are required. Although these atmospheres may explain the evidence for water, a lower (probably less than 1 bar) \ce{CO2} is more likely for the lack of widespread carbonates on Mars's surface \citep{hu_tracing_2015, edwards_carbon_2015}. Similarly, we place an upper limit on \ce{CO2} of about 1.5 bar at 3.8 Ga. In the existing \ce{CO2}-\ce{H2} atmosphere climate models, a 1.5 bar or smaller \ce{CO2} atmosphere produces barely enough warming to produce melting consistent with the geologic evidence, and requires high variability of reduced gas source fluxes \citep{wordsworth_coupled_2021} or over 10\% \ce{H2} \citep{ramirez_warming_2014}. However, the large \ce{N2} abundance indicated by our model results may compensate for the lower \ce{CO2} abundance and generate additional warming. A substantial \ce{H2} abundance combined with the \ce{CO2}-\ce{N2} atmosphere would help explain the evidence for sustained surface conditions for liquid-water \citep[e.g.,][]{carr_oceans_2003,di_achille_ancient_2010,grotzinger_habitable_2014}. A detailed study of the climate and \ce{H2} sources in this scenario is warranted and forthcoming (Adams et al., in prep.).

\subsection{Comparison to Other Modeling Studies and Experimental Data}

The presence of a thick \ce{CO2} atmosphere on ancient Mars implies carbonate deposition occurred primarily in open-water systems. Our models indicate that deposition in open water systems (OWS) is consistent with the explored range of 300-1400 mbar \ce{CO2} deposited in carbonates; however, deposition entirely in shallow subsurface aquifers (SSA) can only occur if less than 100 mbar \ce{CO2} is deposited. This is due to the difference in \ce{CO2} fractionation in these two mechanisms. So, if there was a thick \ce{CO2} atmosphere on ancient Mars, carbonate deposition must have occurred predominantly in OWS in order to explain the modern $\delta^{13}$C value. This finding is consistent with a previous isotopic evolution model that focused solely on carbon \citep{hu_tracing_2015}. A potentially large deep crustal reservoir may contain the volume of carbonates required to explain a \ce{CO2}-rich atmosphere \citep{michalski_deep_2010}; however, directly observed near-surface carbonate deposits are scarce on Mars, where the largest exposure is found in Nili Fossae and contains up to 12 mbar \ce{CO2} \citep{edwards_carbon_2015}. If this indeed indicates a low global carbonate mass on Mars, then it implies that there was a thin \ce{CO2} atmosphere with carbonate deposition occurring in both SSA and OWS. This implication is reversible: evidence for SSA deposition would imply a low global carbonate mass and a \ce{CO2}-poor ancient atmosphere. Our results are consistent with other paleo-pressure estimates \citep{kite_low_2014,kurokawa_lower_2018}, but none of these provide sufficient constraints to narrow the uncertainty described above. Thus, measuring $\delta^{13}$C in Martian carbonate samples (e.g., via Mars Sample Return) could help constrain the volume of surface water present on ancient Mars, the ancient \ce{pCO2}, and the subsequent atmospheric evolution. 


Furthermore, early carbonate deposition (which we parameterize) depends on the ancient atmospheric size and composition (which we predict). In the SSA scenario, we predict low atmospheric \ce{CO2} (up to $\sim$100 mbar) at the time of peak deposition, which is consistent with the assumption that there was no surface liquid water for carbonate deposition to occur in. On the other hand, carbonate deposition in the OWS scenario requires that there was a significant amount of surface liquid water until 3.5 Ga. The greenhouse effect supplied by the atmosphere must thus be large enough to support that water. In our model runs, the OWS scenarios permit high atmospheric \ce{CO2} when large amounts of carbonates are deposited; when combined with an additional greenhouse gas like \ce{H2}, this atmosphere can plausibly generate the warm temperatures required to support liquid water and be consistent with our parameterization. However, the OWS scenarios with low amounts of carbonate deposition (e.g., 300 mbar) require low atmospheric \ce{CO2}; it is likely that these atmospheres could not support liquid water without implausibly high abundances of a greenhouse gas like \ce{H2}. Thus, scenarios for Mars's evolution where there was abundant liquid water but little open water carbonate deposition may be inconsistent. This indicates that our model predictions with more carbonate deposition in open water, and thus higher atmospheric \ce{CO2}, are more plausible than those with less carbonate deposition in open water.


The evolutionary pathways of $\delta^{13}$C that we predict provide context to interpret recently measured $\delta^{13}$C values in Martian soil \citep{house_depleted_2022}. Sample Analysis at Mars (SAM) aboard the Curiosity rover has measured the $\delta^{13}$C value in methane for 24 soil samples at Gale Crater. They found a wide range of values, from $-137 \pm 8\permil$ to $+22 \pm 10\permil$. In our model solutions, atmospheric $\delta^{13}$C in \ce{CO2} varies from $+46\permil$ at the present-day to as low as $-50\permil$ in the Hesperian (Figure \ref{fig:baseline_sample}e). \citet{house_depleted_2022} propose several mechanisms that could explain the very low $\delta^{13}$C measurements that depend on the background value in atmospheric \ce{CO2} including methanogenesis, abiotic reduction of \ce{CO2}, and photochemical reactions. Our results indicate that the process or processes responsible for the very low measured $\delta^{13}$C values must have a significant inherent fractionation effect with respect to the background atmospheric \ce{CO2}; even the lowest $\delta^{13}$C value in our models is still $87\permil$ higher than the lowest measured $\delta^{13}$C.


Our results are broadly consistent with \citet{slipski_argon_2016}, who modeled the argon isotope system. Our model of the argon isotopes followed the general framework of \citet{slipski_argon_2016}, but with several notable differences: our baseline \ce{CO2} sputtering rates are similar in range but different in slope, the crustal production profiles are different, and the atmospheric component of our model only extends to 3.8 Ga whereas theirs extends to 4.4 Ga. Despite these model differences, we find general agreement between the two models. In our baseline MCMC solutions, $f_{\rm ce}$ = 0.4 on average, which is within the 0.31-0.67 range calculated in \citet{slipski_argon_2016} (Figure \ref{fig:baseline_posteriors}). On the other hand, we do not recover the same constraints on the outgassing multiplier in our baseline MCMC: our posterior distributions are pressed against the upper bound of 2, while theirs are constrained to 0.5-1.1. However, in the variant MCMC where atmospheric collapse is ignored, we find the outgassing multiplier is constrained to the range 1-1.7 which is in better agreement because this scenario is more consistent with the \ce{CO2} evolution enforced by \citet{slipski_argon_2016}. The consistency between our models strengthens our conclusions, but the differences highlight the importance of better constraining rates of atmospheric escape and volcanic outgassing.


Our modeling results are consistent with isotopic evolution models that include other noble gases. The nitrogen and noble gas isotopic models presented in \citet{kurokawa_lower_2018} indicate that the atmosphere was greater than 0.5 bar at 4 Ga. This is consistent with our results, as the vast majority of our solutions have an atmosphere greater than 0.5 bar at 3.8 Ga, which would have been subject to loss since 4 Ga (Figure \ref{fig:baseline_comp}d). The neon isotopic models presented in \citet{kurokawa_mars_2021} show evidence for significant recent volcanism on Mars. This strengthens the results we have found, as recent volcanism is important for recreating the modern $\delta^{15}$N, which drives the behavior of all three species we included. The direct inclusion of Ne, Kr, and Xe isotopes in our model would make the constraints derived here more comprehensive. These species were not included here to ensure that the scope of this work is narrow enough to provide a rigorous analysis.


Additional measurements of Mars's past isotopic composition can greatly improve modeling studies like the one presented here. As is shown in a sample of model solutions (Figure \ref{fig:baseline_sample}), the delta values have a wide envelope of possible trajectories that depend on the initial atmospheric composition and model parameter values. Narrowing these envelopes with data will provide even stronger constraints on Mars's atmospheric evolution. Moreover, we are forced to assume delta values for each species at 3.8 Ga based on meteorite analysis and our knowledge of Mars's interior. Future missions such as Mars Sample Return may allow the analysis of samples in contact with Mars's atmosphere from billions of years ago. Isotopic analysis of these samples would be valuable for strengthening the constraints derived from isotopic modeling studies like this.
 
\subsection{Model Validity}

The parameterizations included in our model are consistent with the current best knowledge of planetary processes, but our results imply ancient atmospheric conditions that demand new model studies. A large amount of ancient atmospheric \ce{N2} is indicated by our results; \ce{pN2} is comparable to \ce{pCO2} in baseline solutions and may be larger than \ce{pCO2} in extended range solutions. This introduces uncertainty because our atmospheric escape rates and collapse treatment are derived from a \ce{CO2}-dominated atmosphere. The photochemical escape rate of \ce{N2} may be different in the extended range solutions when the mixing ratio of \ce{N2} is near 1. A detailed photochemical modeling study would be needed to evaluate this, but this uncertainty is mitigated because sputtering is always found to be the dominant loss mechanism in these solutions.

Moreover, our parameterization of atmospheric collapse is subject to uncertainty. First, we base the threshold for collapse on the climate models of \citet{forget_3d_2013}, but other models yield slightly different threshold values \citep[e.g.,][]{soto_martian_2015}. Additionally, the \citet{forget_3d_2013} climate model does not include \ce{N2} as a major component of the atmosphere, which may change the threshold values. Increased study of the thermodynamics of this potential atmosphere is warranted. Second, recent work suggests that Mars's past obliquity may not have been as chaotic as we have assumed \citep{lissauer_obliquity_2012}. Third, it is possible that the atmosphere would not have recovered once it entered a collapsed state due to hysteresis \citep{kurahashi-nakamura_atmospheric_2006}. We do not include this possibility here, but a similar simulation in which the atmosphere is always collapsed can be found in \citet{hu_nitrogen-rich_2022}. Taken together, these considerations indicate that atmospheric collapse throughout Mars history is a wide ranging and uncertain process. Our treatment of atmospheric collapse does not explore the full range of uncertainty, but acts as a representative example for understanding its impact on atmospheric evolution. We include the variant scenario with no atmospheric collapse as an endmember case to highlight this. Future studies will be strengthened by a more comprehensive analysis of atmospheric collapse throughout Mars's history.


We assume a constant concentration of \ce{CO2} in the source magma for volcanic outgassing; however, the \ce{CO2} concentration may have varied with time, as indicated by the range of oxygen fugacities preserved in Martian meteorites and indicated by interior modeling studies \citep[e.g.,][]{warren_siderophile_1996,grott_volcanic_2011}. This is mitigated by the fact that recent volcanic emplacement is orders of magnitude slower than at ancient times, so a changing \ce{CO2} concentration after 3-3.5 Ga is less impactful. Similar studies in the future will be strengthened by coupling mantle redox chemistry explicitly to atmospheric evolution.


Our assumption that the \ce{N2} concentration in Martian source magma is equal to the ``Silicate Earth'' concentration \citep{marty_nitrogen_2003} is subject to uncertainty. On one hand, \ce{N} solubility in magma is dependent on the redox state of the magma, with more reducing conditions allowing more dissolved \ce{N} \citep{boulliung_oxygen_2020}. The early magma ocean on Mars may have been more reducing than Earth's magma ocean \citep{armstrong_deep_2019}, and as discussed above, Mars's mantle since planet solidification may have been more reduced than Earth's mantle. This would lead to a higher \ce{N} concentration in the Martian mantle than in the ``Silicate Earth'' assumption we employ. On the other hand, Earth's mantle \ce{N2} concentration may have been significantly affected by subduction processes that do not occur on Mars \citep{marty_nitrogen_2003, kurokawa_origin_2022}. Thus, it is also possible that the true \ce{N} concentration in the Martian mantle is lower than our ``Silicate Earth'' assumption. Although we do not explicitly vary the model mantle \ce{N} concentration, uncertainty in its value is incorporated into the multipliers of volcanic outgassing and atmospheric escape of nitrogen. Higher values for these multipliers correspond to an enhanced magma \ce{N2} concentration on Mars, and vice versa.


We do not include direct or indirect hydrodynamic escape. Analysis of noble gas isotopes \citep[e.g.,][]{cassata_xenon_2022} indicates that significant hydrodynamic escape of Mars's primary atmosphere ceased before our model time domain. Furthermore, up to 10\% \ce{H2} in the ancient atmosphere should have little effect on the evolution of the other species we model, as hydrodynamic drag from such a small \ce{H2} component should be negligible \citep{zahnle_mass_1986}. Thus, we assume that the minimal hydrodynamic escape during our model time domain is incorporated into the uncertainty of our modeled escape rates via model parameters and multipliers. Reconciling the atmosphere we predict at 3.8 Ga with earlier hydrodynamic escape is an interesting challenge, though. The atmosphere we predict at 3.8 Ga is consistent with modern Mars's escape rates and atmospheric composition; future studies may determine if it is also consistent with the vastly different processes occurring beforehand.


The extended range MCMC simulations further explore the boundary conditions of the model. Many solutions found in the extended range simulations are likely unrealistic because they push the model far from the current best estimates or the empirically constrained ranges. For example, the outgassing multiplier is not likely to be at the upper end of the extended range because we would have the record for much more volcanic activity on Mars. Thus, we feel that the solutions found in the baseline MCMC simulations are probably more realistic for the evolution of Mars's atmosphere. 

\section{Conclusions} \label{sec:conc}

We have presented a comprehensive model that simulates the evolution of \ce{CO2}, \ce{N2}, and \ce{Ar} in the Martian atmosphere from 3.8 billion years ago (Ga) to the present. The model keeps track of both the mass of each species and its isotopic composition through the course of an evolution as the atmosphere is subject to a host of planetary processes. The model includes atmospheric escape (via photochemical processes, pick-up ion sputtering, and direct ion loss), volcanic outgassing, interplanetary dust particle accretion, atmospheric collapse, carbonate formation, nitrate formation, and crustal erosion. There is no predetermined evolution for any atmospheric species. The abundances of the three species are dynamically updated at each timestep according to the planetary processes. The mixing ratios of the three species are calculated accordingly and, in turn, are used to calculate the rate at which planetary processes impact each species for the next timestep. Model runs that recreate the present day abundances and isotopic composition of \ce{CO2}, \ce{N2}, and \ce{Ar} in the Martian atmosphere are considered solutions. We used a Markov-Chain Monte Carlo method to systematically explore the parameter space and identify hundreds of thousands of solutions, derive posterior distributions for parameters, and constrain the composition of Mars's atmosphere at 3.8 Ga.

This analysis yields the following conclusions. (1) Including multiple species in the model is important. Because we jointly analyze \ce{CO2}, \ce{N2}, and \ce{Ar}, constraints originally arising from individual species must now be consistent with all three. For example, the \ce{N2} evolution requires a high volcanic production rate and the \ce{Ar} evolution requires a low sputtering rate. The rates of volcanic production and sputtering affect all three species, so constraints on their values cause downstream constraints on other parameters (e.g., magma \ce{CO2} concentration ($X^C_{\rm mag}$) and upper atmosphere structure ($\Delta z/T$)). Thus, the joint analysis results in new constraints or strengthens existing ones, which improve our understanding of Mars's ancient atmospheric composition and evolution. (2) Carbonate deposition critically determines the ancient \ce{pCO2}. For 0.3-1.4 bar \ce{CO2} in carbonates globally on Mars, we predict 0.3-1.5 bar atmospheric \ce{CO2} at 3.8 Ga. If there was a \ce{CO2}-rich atmosphere on ancient Mars, extensive carbonate deposition in open water on Mars's surface is required to recreate the modern \ce{CO2} abundance and isotopic composition. Conversely, evidence that most carbonate deposition occurred in subsurface aquifers indicates a \ce{CO2}-poor ancient atmosphere. (3) Our models indicate the ancient Martian mantle had oxygen fugacity of IW to IW-1, suggesting a volcanic \ce{H2} flux up to 1.34$\times10^{10}$ cm$^{-2}$ s$^{-1}$ is consistent with a \ce{CO2}-rich atmosphere. This alone cannot support a 5\% \ce{H2} atmosphere required for substantial warming of the surface, so non-volcanic \ce{H2} sources, such as crustal hydration, would be necessary. (4) Our models predict that \ce{N2} was a major component of the atmosphere (0.1-0.5 bar) at 3.8 Ga. This is consistent with the ``dynamical track'' \ce{N2} solutions found in a previous isotopic model \citep{hu_nitrogen-rich_2022}. (5) Our results point to an emerging picture of a \ce{CO2}-\ce{N2}-\ce{H2} greenhouse atmosphere on ancient Mars. We have shown that a \ce{CO2}-rich atmosphere is compatible with hundreds of mbars of \ce{N2} and a potential \ce{H2} component. 


The ancient Martian atmosphere must be consistent with the geologic evidence for surface liquid-water as well as Mars's geochemical evolution. There are many potential atmospheric sizes and compositions that can generate a climate that supports liquid-water on ancient Mars. The evidence from Mars's geochemical evolution can tell us which of these potential atmospheres actually existed. The \ce{CO2}-\ce{N2} atmosphere we propose comes solely from analysis of the geochemical evolution, but may also explain the ancient climate when combined with additional \ce{H2} source fluxes. Thus, the \ce{CO2}-\ce{N2}-\ce{H2} atmosphere is a promising candidate for the ancient atmosphere that explains evidence from both the geochemical evolution and the ancient climate. Studies of the implied climate from this atmosphere and the \textit{in situ} and sample return analysis of the isotopic record will be useful for further refining and testing this prediction. This effort ultimately improves our understanding of ancient Mars: an important, potentially habitable environment elsewhere in the solar system.

\bigskip

We thank Y. L. Yung, B. Ehlmann, and B. Jakosky for helpful discussions. This work was supported by NASA Habitable Worlds grant NNN13D466T, later changed to 80NM0018F0612. The research was carried out at the Jet Propulsion Laboratory, California Institute of Technology, under a contract with the National Aeronautics and Space Administration.

\bigskip

The Matlab source code for the coupled atmospheric evolution model, the associated configuration files used in this study, and the data required to make the figures in the main text are publicly available at Zenodo (\url{https://doi.org/10.5281/zenodo.7600495}).

\bibliography{bibliography}{}
\bibliographystyle{aasjournal}

\end{document}